\documentclass[12pt,a4paper]{article}
\usepackage{graphicx}
\usepackage{subeqnarray}
\begin{document}
%
%
%
%
\def\astrobj#1{#1}
\newenvironment{lefteqnarray}{\arraycolsep=0pt\begin{eqnarray}}
{\end{eqnarray}\protect\aftergroup\ignorespaces}
\newenvironment{lefteqnarray*}{\arraycolsep=0pt\begin{eqnarray*}}
{\end{eqnarray*}\protect\aftergroup\ignorespaces}
\newenvironment{leftsubeqnarray}{\arraycolsep=0pt\begin{subeqnarray}}
{\end{subeqnarray}\protect\aftergroup\ignorespaces}
\newcommand{\diff}{{\rm\,d}}
\newcommand{\pprime}{{\prime\prime}}
\newcommand{\szeta}{\mskip 3mu /\mskip-10mu \zeta}
\newcommand{\FC}{\mskip 0mu {\rm F}\mskip-10mu{\rm C}}
\newcommand{\appleq}{\stackrel{<}{\sim}}
\newcommand{\appgeq}{\stackrel{>}{\sim}}
\newcommand{\Int}{\mathop{\rm Int}\nolimits}
\newcommand{\Nint}{\mathop{\rm Nint}\nolimits}
\newcommand{\arcsinh}{\mathop{\rm arcsinh}\nolimits}
\newcommand{\range}{{\rm -}}
\newcommand{\displayfrac}[2]{\frac{\displaystyle #1}{\displaystyle #2}}
\def\astrobj#1{#1}
%
\title{Gravitational acceleration and tidal effects \\
in spherical-symmetric density profiles}
\author{
{R.~Caimmi}\footnote{
{\it Physics and Astronomy Department, Padua University,
Vicolo Osservatorio 3/2,
I-35122 Padova, Italy}
email: roberto.caimmi@unipd.it~~~
fax: 39-049-8278212}
\phantom{agga}}

%
%
\maketitle
\begin{quotation}
\section*{}
\begin{Large}
\begin{center}

Abstract

\end{center}
\end{Large}
\begin{small}

\noindent\noindent
Pure power-law density profiles, $\rho(r)\propto r^{b-3}$, are classified in
connection with the following reference cases: (i) isodensity, $b=3$,
$\rho=$ const; (ii) isogravity, $b=2$, $g=$ const; (iii) isothermal, $b=1$,
$v=[GM(r)/r]^{1/2}=$ const; (iv) isomass, $b=0$, $M=$ const.   A restricted
number of different families of density profiles including, in addition, cored
power-law, generalized power-law, polytropes, are studied in detail with
regard to both one-component and two-component systems.   Considerable effort
is devoted to the existence of an extremum point (maximum absolute value) in
the gravitational acceleration within the matter distribution.   Predicted
velocity curves are compared to the data inferred from observations.   Tidal
effects on an inner subsystem are investigated and an application is made to
globular clusters within the Galaxy.   To this aim, the tidal radius is
defined by balancing the opposite gravitational forces from the Galaxy and the
selected cluster on a special point of the cluster boundary, lying between
related
centres of mass.   The position of 17 globular clusters with respect to the
stability region, where the tidal radius exceeds the observed radius, is shown
for assigned dark-to-visible mass ratios and
density profiles, among those considered, which are currently used for the
description of galaxies and/or dark matter haloes.

\noindent
{\it keywords - 
cosmology: dark matter - galaxies: structure - globular clusters: general.}
\end{small}
\end{quotation}

\section{Introduction} \label{intro}

Typical galaxies, such as spirals and giant ellipticals, are multi-component
systems mainly substructured as (1) nonbaryonic dark halo; (2) baryonic halo;
(3) disk; (4) bulge; (5) inner accretion disk; (6) central accreting
supermassive black hole, in brief hole.   For sufficiently extended subsystems
i.e. (1)-(4), a description in terms of density profiles enables the
determination of the theoretical circular velocity profile and the comparison
with its empirical counterpart inferred from observations (e.g., Haud and
Einasto 1989).

Strictly speaking, density profiles should relate to equilibrium equations
(e.g., Jaffe, 1983; Hernquist, 1990; Dehnen, 1993; Stone and Ostriker, 2015)
but an acceptable fit to light distribution, via a mass-luminosity relation,
could be sufficient for the description of related subsystems (e.g., De
Vaucouleurs, 1948; Sersic, 1963; Kormendy, 1977).

In most cases, each subsystem is treated separately and the model is a simple
superposition of the various components.   More self-consistent models should
account for the fact that each subsystem ought to readjust under the tidal
action from the remaining subsystems.   On the other hand, a simple
superposition model may be adequate in deriving the general properties of a
mass distribution, provided the number of free parameters remains reasonably
low (e.g., Carignan 1985).

Though a quantitative (i.e. with regard to parameter values) classification of
density profiles for one-component (e.g., Zhao 1996; An and Zhao, 2013) and
two-component (e.g., Ciotti and Pellegrini 1992; Ciotti, 1999) is long dating,
still less effort has been devoted to a qualitative (i.e. with regard to
intrinsic properties) classification.   To this aim, the current paper takes
into consideration a restricted number of density profiles, part obeying an
equilibrium equation and part being purely descriptive.
For sake of simplicity, attention shall be restricted to spherical-symmetric
configurations, even if the formulation maintains for homeoidally striated
ellipsoidal configurations as far as radial properties are concerned (e.g.,
Caimmi 1993, 2003; Caimmi and Marmo 2003).   To emphasize the above mentioned
property, spherical-symmetric density profiles shall be quoted henceforth as
homeoidally striated spherical density profiles. 

Pure power-law density profiles, $\rho(r)\propto r^{b-3}$, $0\le b\le3$, are
used to perform a qualitative classification characterized by the following
reference cases: (i) isodensity, $b=3$,
$\rho=$ const; (ii) isogravity, $b=2$, $g=$ const; (iii) isothermal, $b=1$,
$v=[GM(r)/r]^{1/2}=$ const; (iv) isomass, $b=0$, $M=$ const.

In the last case,
$\rho(r)=m_0/r^3$, where $m_0$ is a proportionality constant dimensioned as a
mass which, in addition, has to be infinitesimal of higher order with respect
to $1/\log r$ to avoid a logarithmic divergence in the mass, $M(r)$.   More
specifically, $m_0$ has to be infinitesimal of equal order with respect to
$r^3$ to ensure a finite nonzero mass at the centre, $0<M(0)=M(r)<+\infty$.

A selected density
profile can be classified as pseudo isodense, pseudo isogravity, pseudo
isothermal, pseudo isomass, according if it is sufficiently close to the
appropriate reference case.   In particular, density profiles where $3\ge b>2$
in the inner regions and $2>b\ge0$ in the outer regions, could exhibit a
nonmonotonic gravitational acceleration with the occurrence of an extremum
point (maximum absolute value) inside the domain.   Accordingly, further
classification of
density profiles should include the presence or the absence of a maximum in
$\vert g\vert$ within a selected matter distribution.

Additional investigation, conceived as an application of the general theory,
can be devoted to the stability of globular clusters against the tidal action
of the Galaxy via a simple formulation of tidal radius, involving balance of
opposite gravitational forces exerted from a selected cluster and the Galaxy
on a special point on the cluster surface lying between related centres of
mass (e.g., Von Hoerner 1958; Brosche et al. 1999).   For a fixed density
profile, clusters where the tidal radius exceeds
the observed radius are considered bound, if otherwise (partially) unbound.

The position
of globular clusters can be plotted together with the stability region, and
the number of bound and (partially) unbound globular clusters can be
determined for
selected Galactic density profiles.   Finally, Galactic density profiles can
be constrained according if globular clusters showing no sign of tidal tails
or tidal streams lie within the instability region and vice versa.
Though a more accurate definition of tidal radius should be used to infer
quantitative conclusions, still the trend is expected to be similar and the
validity of the method is left unchanged.

The current paper is structured in the following way.  One-component,
homeoidally striated spherical density profiles are described in Section
\ref{hssp},
where further attention is devoted to a few families of density profiles,
namely (a) pure power-law; (b) cored power-law; (c) polytropes; (d) Plummer
(1911); (e) Hernquist (1990); (f) Begeman et al. (1991); (g) Spano et al.
(2008); (h) Burkert (1995).   Two-component, concentric, homeoidally striated
spherical density profiles are described in Section \ref{2cdp},
where further attention is devoted to a few combinations of density profiles, 
namely (i) Plummer-Begeman et al.; (j) Hernquist-Begeman et
al; (k) Plummer-Spano et al.; (l) Hernquist-Spano et al.  Tidal effects on
embedded sybsystems are analysed in Section \ref{app}, where the tidal radius
is defined via gravitational balance on a special point and an application to
globular clusters within galaxies is performed,
with regard to density profiles currently used for the description of
galaxies and/or dark matter haloes, among those considered.   The results are
discussed in Section \ref{disc}, where the particularization to a sample of
Galactic
globular clusters is also shown.   Finally, the conclusion is drawn in
Section \ref{conc}.  Further details on specific arguments are shown in the
Appendix.

\section{Homeoidally striated spherical density profiles}
\label{hssp}

With respect to a reference frame where the origin coincides with the centre
of mass, let $\rho(r)$ be a generic, homeoidally striated spherical density
profile.
Accordingly, the equipotential surfaces are concentric spheres centered on the
origin. The mass, the squared circular velocity, and the gravitational
acceleration profile, respectively, are:
\begin{lefteqnarray}
\label{eq:Mr}
&& M(r)=4\pi\int_0^r\rho(r)r^2\diff r~~; \\
\label{eq:v2}
&& v^2(r)=\frac{GM(r)}r~~; \\
\label{eq:g}
&& g(r)=-\frac{GM(r)}{r^2}~~;
\end{lefteqnarray}
and the local slope of the gravitational acceleration profile reads:
\begin{lefteqnarray}
\label{eq:dgdr}
&& \frac{\diff g}{\diff r}=G\left[\frac{2M(r)}{r^3}-\frac1{r^2}\frac{\diff M}
{\diff r}\right]~~; \\
\label{eq:dMdr}
&& \frac{\diff M}{\diff r}=4\pi\rho(r)r^2~~;
\end{lefteqnarray}
where $G$ is the constant of gravitation.

Extremum points of the gravitational acceleration profile must fulfill the
condition, $\diff g/\diff r=0$, or:
\begin{equation}
\label{eq:estg}
M(r)=2\pi r^3\rho(r)~~;
\end{equation}
which, in terms of the global density, translates into:
\begin{equation}
\label{eq:rhor}
\bar{\rho}(r)=\frac3{4\pi}\frac{M(r)}{r^3}=\frac32\rho(r)~~;
\end{equation}
where the radial coordinate, satisfying Eq.\,(\ref{eq:rhor}), is the
extremum point of the gravitational acceleration profile i.e. the first
derivative of the gravitational potential.   Keeping in mind
no gravitational force is exerted at the origin, $r=0$, the
extremum point has to be a minimum (maximum in absolute value).
For an extension to homeoidally striated ellipsoidal density profiles, an
interested reader is addressed to Appendix \ref{a:elli}.

In dimensionless coordinates, a generic density profile reads (e.g., Caimmi
and Marmo 2003):
\begin{lefteqnarray}
\label{eq:rho}
&& \rho(r)=\rho^\dagger f(\xi)~~;\qquad0\le\xi\le\Xi~~; \\
\label{eq:csif}
&& \xi=\frac r{r^\dagger}~~;\qquad\Xi=\frac R{r^\dagger}~~;\qquad f(1)=1~~;
\end{lefteqnarray}
where $\rho^\dagger$, $r^\dagger$, are a scaling density and a scaling radius,
respectively, $R$ is the truncation radius, and $\rho^\dagger=\rho(r^\dagger)$
via Eq.\,(\ref{eq:csif}).   The logarithmic slope at the dimensionless scaling
radius, hereafter quoted as the scaling logarithmic slope, is
$(\diff\log f/\diff\log\xi)_{\log\xi=0}$ (Caimmi et al. 2005).  It shall be
intended any density profile may be extended outside the truncation radius,
putting $\rho(r)=0$, $r>R$.

The related mass profile reads (e.g., Caimmi and Marmo 2003):
\begin{lefteqnarray}
\label{eq:M}
&& M(r)=M^\dagger\nu_{\rm mas}(\xi)~~; \\
\label{eq:Mc}
&& M^\dagger=\frac{4\pi}3\rho^\dagger(r^\dagger)^3~~;\qquad\nu_{\rm mas}(\xi)=
\frac32\left[\int_0^\xi F(\xi)\diff\xi-\xi F(\xi)\right]~~;
\end{lefteqnarray}
where $M^\dagger$ is a scaling mass, $\nu_{\rm mas}$ is a profile factor i.e.
depending on $\xi$ only, and the integrand is defined as:
\begin{equation}
\label{eq:F}
F(\xi)=2\int_\xi^\Xi f(\xi)\xi\diff\xi~~;\qquad F(\Xi)=0~~;\qquad\frac
{\diff F}{\diff\xi}=-2\xi f(\xi)~~;
\end{equation}
for further details, an interested reader is addressed to the parent paper
(Roberts 1962).

The related global density, by definition, is:
\begin{equation}
\label{eq:rhom}
\bar{\rho}(r)=\frac3{4\pi}\frac{M(r)}{r^3}=\frac3{4\pi}\frac{M^\dagger}
{(r^\dagger)^3}\frac{\nu_{\rm mas}(\xi)}{\xi^3}=\rho^\dagger\frac
{\nu_{\rm mas}(\xi)}{\xi^3}~~;
\end{equation}
and the ratio of global to local density reads:
\begin{equation}
\label{eq:rmrl}
\frac{\bar{\rho}(r)}{\rho(r)}=\frac{\nu_{\rm mas}(\xi)}{\xi^3f(\xi)}~~;
\end{equation}
the substitution of Eqs.\,(\ref{eq:csif}) and (\ref{eq:M}) into (\ref{eq:v2})
and (\ref{eq:g}) yields the squared circular velocity and gravitational
acceleration profiles in dimensionless coordinates, as:
\begin{lefteqnarray}
\label{eq:v2csi}
&& v^2(r)=(v^\dagger)^2\frac{\nu_{\rm mas}(\xi)}\xi~~; \\
\label{eq:v2c}
&& (v^\dagger)^2=\frac{GM^\dagger}{r^\dagger}~~; \\
\label{eq:gcsi}
&& g(r)=g^\dagger\frac{\nu_{\rm mas}(\xi)}{\xi^2}~~; \\
\label{eq:gc}
&& g^\dagger=-\frac{GM^\dagger}{(r^\dagger)^2}~~;
\end{lefteqnarray}
where $v^\dagger$ and $g^\dagger$ are a scaling circular velocity and a
scaling gravitational acceleration, respectively.

The combination of Eqs.\,(\ref{eq:M}), (\ref{eq:Mc}),
(\ref{eq:v2csi})-(\ref{eq:gc}), yields:
\begin{equation}
\label{eq:numvg}
\nu_{\rm mas}(\xi)=\frac{M(r)}{M^\dagger}=\left[\frac{v(r)}{v^\dagger}\right]^
2\xi=\frac{g(r)}{g^\dagger}\xi^2~~;
\end{equation}
and the combination of Eqs.\,(\ref{eq:rhor}) and (\ref{eq:rmrl}) yields:
\begin{equation}
\label{eq:gest}
\frac{\nu_{\rm mas}(\xi)}{\xi^3f(\xi)}=\frac32~~;
\end{equation}
the solution of which is the extremum point of the gravitational acceleration
profile in dimensionless coordinates.

Let the isodensity surface where the gravitational acceleration attains the
extremum point be defined as effective surface of the system, and the
related radius as effective radius.   In absence of an extremum point,
the effective radius necessarily takes place on the boundary of the
domain, either $r_{\rm eff}=0$ or $r_{\rm eff}=R$, in dimensionless coordinates
either $\xi_{\rm eff}=0$ or $\xi_{\rm eff}=\Xi$.

The particularization of the above results to some special families of density
profiles shall be performed in the following subsections.

\subsection{Pure power-law density profiles}
\label{pola}

Pure power-law density profiles are defined as (e.g., Caimmi 2008):
\begin{equation}
\label{eq:fpl}
f(\xi)=\xi^{b-3}~~;\qquad0\le\xi\le\Xi~~;\qquad0\le b\le3~~;
\end{equation}
where $b>3$ implies increasing density for increasing radial coordinate, and
$b<0$ implies infinite central mass.   The scaling logarithmic slope is:
\begin{equation}
\label{eq:ppl}
\left(\frac{\diff\log f}{\diff\log\xi}\right)_{\log\xi=0}=b-3~~;
\end{equation}
which is constant in the case under consideration.

The substitution of Eq.\,(\ref{eq:fpl}) into (\ref{eq:M})-(\ref{eq:gc})
yields after some algebra:
\begin{lefteqnarray}
\label{eq:Fpl}
&& F(\xi)=\frac2{b-1}(\Xi^{b-1}-\xi^{b-1})~~; \\
\label{eq:numpl}
&& \nu_{\rm mas}(\xi)=\frac3b\xi^b~~; \\
\label{eq:rompl}
&& \frac{\bar{\rho}(r)}{\rho^\dagger}=\frac3b\xi^{b-3}~~; \\
\label{eq:rhorpl}
&& \frac{\bar{\rho}(r)}{\rho(r)}=\frac3b~~; \\
\label{eq:v2pl}
&& \frac{v^2(r)}{(v^\dagger)^2}=\frac3b\xi^{b-1}~~; \\
\label{eq:gpl}
&& \frac g{g^\dagger}=\frac3b\xi^{b-2}~~;
\end{lefteqnarray}
where $b=3$ corresponds to the isodensity ($\rho=$ const) sphere, $b=2$ to the
isogravity ($g=$ const) sphere, $b=1$ to the isothermal
($v^2=GM(r)/r=$ const) sphere, $b=0$ to the isomass ($M=$ const) sphere,
according to Eqs.\,(\ref{eq:rho}) and (\ref{eq:fpl}).

In the last case, $b=0$ via
Eq.\,(\ref{eq:rhorpl}) implies either infinite mass for finite local density,
or finite mass for null local density outside the centre of mass.
Restricting to
the latter alternative (isomass sphere), it can be seen that $\rho^\dagger\to
0$, $M^\dagger\to0$, $\nu_{\rm mas}(\xi)\to+\infty$, $M(\xi)=M(\Xi)=M$, and
the density profile represents a mass point surrounded by a massless
atmosphere i.e. a Roche system.

According to Eq.\,(\ref{eq:gpl}), power-law density profiles exhibit no
(absolute) extremum point of the gravitational acceleration.

\subsection{Cored power-law density profiles}
\label{cpla}

Aiming to eliminate the central density cusp exhibited by pure power-law
density
profiles, cored power-law density profiles are defined as (e.g., Secco 2005;
Caimmi 2008):
\begin{equation}
\label{eq:fcl}
f(\xi)=\cases{
1~~; & $0\le\xi\le1~~;$ \cr
\xi^{b-3}~~; & $1\le\xi\le\Xi~~;$ \cr}
\end{equation}
where the inner part of related power-law density profile is replaced by a
homogeneous sphere of density, $\rho^\dagger=\rho(r^\dagger)$.
The scaling logarithmic slope is discontinuous with a null value at $\log\xi
\to0^-$ and a value expressed by Eq.\,(\ref{eq:ppl}) at $\log\xi\to0^+$.

The substitution of Eq.\,(\ref{eq:fcl}) into (\ref{eq:M})-(\ref{eq:gc})
yields after some algebra:
\begin{lefteqnarray}
\label{eq:Fcl}
&& F(\xi)=\cases{
(1-\xi^2)+\frac2{b-1}(\Xi^{b-1}-1)~~; & $0\le\xi\le1~~;$ \cr
\frac2{b-1}(\Xi^{b-1}-\xi^{b-1})~~; & $1\le\xi\le\Xi~~;$ \cr} \\
\label{eq:numcl}
&& \nu_{\rm mas}(\xi)=\cases{
\xi^3~~; & $0\le\xi\le1~~;$ \cr
1+\frac3b(\xi^b-1)~~; & $1\le\xi\le\Xi~~;$ \cr} \\
\label{eq:romcl}
&& \frac{\bar{\rho}(r)}{\rho^\dagger}=\cases{
1~~; & $0\le\xi\le1~~;$ \cr
\frac1{\xi^3}+\frac3b\frac{\xi^b-1}{\xi^3}~~; & $1\le\xi\le\Xi~~;$ \cr}  \\
\label{eq:rhorcl}
&& \frac{\bar{\rho}(r)}{\rho(r)}=\cases{
1~~; & $0\le\xi\le1~~;$ \cr
\frac1{\xi^b}+\frac3b\frac{\xi^b-1}{\xi^b}~~; & $1\le\xi\le\Xi~~;$ \cr} \\
\label{eq:v2cl}
&& \frac{v^2(r)}{(v^\dagger)^2}=\cases{
\xi^2~~; & $0\le\xi\le1~~;$ \cr
\frac1\xi+\frac3b\frac{\xi^b-1}\xi~~; & $1\le\xi\le\Xi~~;$ \cr}  \\
\label{eq:gcl}
&& \frac g{g^\dagger}=\cases{
\xi~~; & $0\le\xi\le1~~;$ \cr
\frac1{\xi^2}+\frac3b\frac{\xi^b-1}{\xi^2}~~; & $1\le\xi\le\Xi~~;$ \cr} 
\end{lefteqnarray}
where $b=3$ corresponds to the isodensity ($\rho=$const) sphere.
The substitution of Eqs.\,(\ref{eq:fcl}) and (\ref{eq:numcl}) into
(\ref{eq:gest}) yields after some algebra:
\begin{equation}
\label{eq:gecl}
\frac3b\left[1-\frac{3-b}3\xi^{-b}\right]=\frac32~~;
\end{equation}
and the extremum point of the gravitational acceleration is defined by the
solution of Eq.\,(\ref{eq:gecl}) as:
\begin{equation}
\label{eq:sgcl}
\xi=\left(\frac23\,\frac{3-b}{2-b}\right)^{1/b}~~;
\end{equation}
where $\xi\ge1$ in that no extremum point exists for the inner homogeneous
sphere,
$0\le\xi\le1$, according to Eq.\,(\ref{eq:gpl}), and, on the other hand,
$\xi\le\Xi$, according to Eq.\,(\ref{eq:fcl}), which implies the following:
\begin{lefteqnarray}
\label{eq:cocl}
&& 1\le\phi(b)=\left(\frac23\,\frac{3-b}{2-b}\right)^{1/b}\le\Xi~~; \\
\label{eq:licl}
&& \lim_{b\to0}\left(\frac23\,\frac{3-b}{2-b}\right)^{1/b}=\exp\left(\frac16
\right)~~;
\end{lefteqnarray}
where $\phi(b)$ is defined within the domain, $0\le b<2$, $b=3$, as powers
with real exponents must necessarily exhibit nonnegative basis, and is
monotonically increasing.

In conclusion, an extremum point is attained by the gravitational
acceleration for sufficiently steep $(0\le b<2)$ cored power-law density
profiles, provided Eq.\,(\ref{eq:cocl}) is satisfied.   In particular,
$\Xi\ge\exp(1/6)$ for $b=0$ i.e. a generalized Roche system.

\subsection{Polytropic density profiles}
\label{poly}

Polytropic spheres obey the Lane-Emden equation (e.g., Chandrasekhar 1939,
Chap.\,IV):
\begin{lefteqnarray}
\label{eq:LE}
&& \frac1{\xi_{\rm LE}^2}\frac\diff{\diff\xi_{\rm LE}}\left(\xi_{\rm LE}^2
\frac{\diff\theta}{\diff\xi_{\rm LE}}\right)=-\theta^n~~; \\
\label{eq:bct}
&& \theta(0)=1~~;\qquad\theta(\Xi_{\rm LE})=0~~;\qquad
\left(\frac{\diff\theta}{\diff\xi_{\rm LE}}\right)_0=0~~; \\
\label{eq:sd0}
&& \theta(\xi_{\rm LE})=1-\frac16\xi_{\rm LE}^2+\frac n{120}\xi_{\rm LE}^4-
...~~;\qquad0\le\xi_{\rm LE}<1~~;
\end{lefteqnarray}
where $n$ is the polytropic index and the dimensionless radial coordinate,
$\xi_{\rm LE}$, and the dimensionless density, $\theta^n$, are related to
their dimensional counterparts as:
\begin{lefteqnarray}
\label{eq:alfa}
&& r=\alpha_{\rm LE}\xi_{\rm LE}~~;\qquad R=\alpha_{\rm LE}\Xi_{\rm LE}~~;
\qquad\alpha_{\rm LE}=\left[\frac{(n+1)K\lambda_{\rm LE}^{1+1/n}}{4\pi G
\lambda_{\rm LE}^2}\right]^{1/2}~~;\qquad \\
\label{eq:rhoLE}
&& \rho(r)=\lambda_{\rm LE}\theta^n(\xi_{\rm LE})~~;\qquad\rho(0)=
\lambda_{\rm LE}~~;\qquad\rho(R)=0~~;
\end{lefteqnarray}
where $\lambda_{\rm LE}$ is the central density, $K\lambda_{\rm LE}^{1+1/n}$
the central pressure, and $\alpha_{\rm LE}$ the polytropic scaling radius.
Polytropic density profiles lie between the limiting cases of homogeneous
configurations $(n=0)$ and Roche systems or Plummer systems $(n=5)$.

The relations between current and polytropic dimensionless density and radial
coordinate, may be deduced by comparison of Eqs.\,(\ref{eq:rho}) and
(\ref{eq:csif}) with (\ref{eq:rhoLE}) and (\ref{eq:alfa}), respectively.
The result is:
\begin{lefteqnarray}
\label{eq:ftheta}
&& f(\xi)=\frac{\lambda_{\rm LE}}{\rho^\dagger}\theta^n(\xi_{\rm LE})~~; \\
\label{eq:ccLE}
&& \xi=\frac{\alpha_{\rm LE}}{r^\dagger}\xi_{\rm LE}~~;\qquad\Xi=\frac
{\alpha_{\rm LE}}{r^\dagger}\Xi_{\rm LE}~~;
\end{lefteqnarray}
and the boundary condition, $f(1)=1$, translates into:
\begin{equation}
\label{eq:bcth}
\theta\left(\frac{r^\dagger}{\alpha_{\rm LE}}\right)=\left(\frac{\rho^\dagger}
{\lambda_{\rm LE}}\right)^{1/n}~~;
\end{equation}
where, in particular:
\begin{equation}
\label{eq:bct0}
\lim_{n\to0}\left(\frac{\rho^\dagger}{\lambda_{\rm LE}}\right)^{1/n}=1-\frac16
\left(\frac{r^\dagger}{\alpha_{\rm LE}}\right)^2~~;
\end{equation}
for homogeneous configurations, using next Eq.\,(\ref{eq:th0}).

As the Lane-Emden equation, Eq.\,(\ref{eq:LE}), and related solutions, have
widely been studied in literature (e.g., Chandrasekhar 1939, Chap.\,IV; Horedt
2004),
the quantities of interest for polytropic density profiles shall be expressed
in terms of the dimensionless density, $\theta^n$, and the dimensionless
radial coordinate, $\xi_{\rm LE}$.

Accordingly, the scaling logarithmic slope is:
\begin{lefteqnarray}
\label{eq:pLE}
&& \left(\frac{\diff\log f}{\diff\log\xi}\right)_{\log\xi=0}=
n\left(\frac{\diff\log\theta}{\diff\log\xi_{\rm LE}}\frac{\diff\log\xi_
{\rm LE}}{\diff\log\xi}\right)_{\log\xi=0} \nonumber \\
&& \phantom{\left(\frac{\diff\log f}{\diff\log\xi}\right)_{\log\xi=0}}=
n\left(\frac{\diff\log\theta}{\diff
\log\xi_{\rm LE}}\right)_{\log\xi_{\rm LE}=\log(r^\dagger/\alpha_{\rm LE})}~~;
\end{lefteqnarray}
and Eqs.\,(\ref{eq:M}), (\ref{eq:Mc}), (\ref{eq:rhom}), (\ref{eq:rmrl}),
(\ref{eq:v2csi}), (\ref{eq:gcsi}) and (\ref{eq:gest}), take the equivalent
form (e.g., Chandrasekhar 1939, Chap.\,IV, \S 6):
\begin{lefteqnarray}
\label{eq:Mpo}
&& M(r)=-M_{\rm LE}\xi_{\rm LE}^2\frac{\diff\theta}{\xi_{\rm LE}}~~; \\
\label{eq:MLE}
&& M_{\rm LE}=4\pi\alpha_{\rm LE}^3\lambda_{\rm LE}~~; \\
\label{eq:rhomp}
&& \frac{\bar{\rho}(r)}{\lambda_{\rm LE}}=-\frac3{\xi_{\rm LE}}\frac{\diff
\theta}{\diff\xi_{\rm LE}}~~; \\
\label{eq:rmrp}
&& \frac{\bar{\rho}(r)}{\rho(r)}=-\frac3{\xi_{\rm LE}}\frac{\diff\theta/\diff
\xi_{\rm LE}}{\theta^n(\xi_{\rm LE})}~~; \\
\label{eq:v2p}
&& \left[\frac{v(r)}{v_{\rm LE}}\right]^2=\frac{GM(r)}{v_{\rm LE}^2r}=
-\xi_{\rm LE}\frac{\diff\theta}{\diff\xi_{\rm LE}}~~; \\
\label{eq:vLE}
&& v_{\rm LE}^2=\frac{G M_{\rm LE}}{\alpha_{\rm LE}}~~; \\
\label{eq:gp}
&& \frac{g(r)}{g_{\rm LE}}=-\frac{GM(r)}{g_{\rm LE}\,r^2}=-\frac{\diff\theta}
{\diff\xi_{\rm LE}}~~; \\
\label{eq:gLE}
&& g_{\rm LE}=-\frac{G M_{\rm LE}}{\alpha_{\rm LE}^2}~~; \\
\label{eq:gesp}
&& \frac1{\xi_{\rm LE}}\frac{\diff\theta}{\diff\xi_{\rm LE}}=-\frac12\theta^n
(\xi_{\rm LE})~~;
\end{lefteqnarray}
where the last result follows from Eq.\,(\ref{eq:bct}), implying the
existence of a solution to Eq.\,(\ref{eq:gesp}) for $n>0$, as $\theta$ is
continuous together with its first and second derivates via
Eq.\,(\ref{eq:LE}).

The substitution of Eq.\,(\ref{eq:LE}) into (\ref{eq:gesp}) yields after some
algebra:
\begin{equation}
\label{eq:gsp2}
\frac{\diff^2\theta}{\diff\xi_{\rm LE}^2}=0~~;
\end{equation}
which shows that, for polytropic density profiles, the extremum point of the
gravitational acceleration coincides with the extremum point of the first
derivative of $\theta$, as expected, the gravitational potential inside
polytropic spheres being proportional to $\theta$ (e.g., Chandrasekhar 1939,
Chap.\,IV, \S 7).

In the special cases, $n=0,1,5$, Eq.\,(\ref{eq:LE}) can analytically be
integrated.   The result is (e.g., Chandrasekhar 1939, Chap.\,IV, \S 4):
\begin{lefteqnarray}
\label{eq:th0}
&& \theta(\xi_{\rm LE})=1-\frac16\xi_{\rm LE}^2~~;\qquad\Xi_{\rm LE}=\sqrt{6}
~~;\qquad n=0~~; \\
\label{eq:th1}
&& \theta(\xi_{\rm LE})=\frac{\sin\xi_{\rm LE}}{\xi_{\rm LE}}~~;\qquad
\Xi_{\rm LE}=\pi~~;\qquad n=1~~; \\
\label{eq:th5}
&& \theta(\xi_{\rm LE})=\left(1+\frac13\xi_{\rm LE}^2\right)^{-1/2}~~;\qquad
\Xi_{\rm LE}\to+\infty~~;\qquad n=5~~;
\end{lefteqnarray}
accordingly, Eq.\,(\ref{eq:gesp}) has no solution for $n=0$, as expected for
homogeneous configurations, and:
\begin{equation}
\label{eq:gep1}
\frac{\xi_{\rm LE}\cos\xi_{\rm LE}-\sin\xi_{\rm LE}}{\xi_{\rm LE}^2}=-\frac12
\sin\xi_{\rm LE}~~;
\end{equation}
for $n=1$, which has a solution, $\xi_{\rm LE}\approx2.081\,575\,99$, and:
\begin{equation}
\label{eq:gep5}
-\frac12\left(1+\frac13\xi_{\rm LE}^2\right)^{-3/2}\frac23\xi_{\rm LE}=-\frac
12\xi_{\rm LE}\left(1+\frac13\xi_{\rm LE}^2\right)^{-5/2}~~;
\end{equation}
for $n=5$, which has a solution, $\xi_{\rm LE}=\sqrt{3/2}\approx
1.224\,744\,871$.

In conclusion, an extremum point is exhibited by the gravitational
acceleration for inhomogeneous polytropic density profiles
$(0<n_{\rm min}\le n\le5)$, where
the solution of related transcendental equation can readily be determined in
the special cases, $n=1,5$, via Eqs.\,(\ref{eq:gep1}), (\ref{eq:gep5}),
respectively.  The threshold, $n_{\rm min}$, $0<n_{\rm min}<1$, is defined by
equating the solution of Eq.\,(\ref{eq:gesp}) to the dimensionless radius,
$\xi_{\rm LE}(n_{\rm min})=\Xi_{\rm LE}(n_{\rm min})$, as $n<n_{\rm min}$
would imply $\xi_{\rm LE}(n_{\rm min})>\Xi_{\rm LE}(n_{\rm min})$, which lies
outside the domain, $0\le\xi_{\rm LE}\le\Xi_{\rm LE}$.

\subsection{Plummer density profiles}
\label{plum}

Plummer density profiles (Plummer 1911), hereafter quoted as P density
profiles, are nothing but $n=5$ polytropes represented by
Eqs.\,(\ref{eq:rho}), (\ref{eq:csif}), via (\ref{eq:ftheta})-(\ref{eq:bcth}),
where parameters are interrelated as (Caimmi and Valentinuzzi 2008):
\begin{equation}
\label{eq:PPLE}
\xi=\frac{\xi_{\rm LE}}{\sqrt{3}}~~;\qquad r^\dagger=\sqrt{3}\,\alpha_{\rm LE}
~~;\qquad\rho^\dagger=\frac{\lambda_{\rm LE}}{2^{5/2}}~~;
\end{equation}
the result is (Caimmi and Valentinuzzi 2008):
\begin{equation}
\label{eq:fP}
f(\xi)=\frac{2^{5/2}}{(1+\xi^2)^{5/2}}~~;
\end{equation}
where the scaling logarithmic slope reads:
\begin{equation}
\label{eq:pP}
\left(\frac{\diff\log f}{\diff\log\xi}\right)_{\log\xi=0}=-\frac52~~;
\end{equation}
and the substitution of Eq.\,(\ref{eq:fP}) into (\ref{eq:M})-(\ref{eq:gc})
yields after some algebra:
\begin{lefteqnarray}
\label{eq:FP}
&& F(\xi)=\frac{2^{7/2}}3\left[\frac1{(1+\xi^2)^{3/2}}-\frac1{(1+\Xi^2)^{3/2}}
\right]~~; \\
\label{eq:numP}
&& \nu_{\rm mas}(\xi)=\frac{2^{5/2}\xi^3}{(1+\xi^2)^{3/2}}~~; \\
\label{eq:rhomP}
&& \frac{\bar{\rho}(r)}{\rho^\dagger}=\frac{2^{5/2}}{(1+\xi^2)^{3/2}}~~; \\
\label{eq:rmrP}
&& \frac{\bar{\rho}(r)}{\rho(r)}=1+\xi^2~~; \\
\label{eq:v2P}
&& \left[\frac{v(r)}{v^\dagger}\right]^2=\frac{2^{5/2}\xi^2}{(1+\xi^2)^{3/2}}
~~; \\
\label{eq:gcsiP}
&& \frac{g(r)}{g^\dagger}=\frac{2^{5/2}\xi}{(1+\xi^2)^{3/2}}~~;
\end{lefteqnarray}
finally, the particularization of Eq.\,(\ref{eq:gest}) to the case under
consideration reads:
\begin{equation}
\label{eq:gestP}
1+\xi^2=\frac32~~;
\end{equation}
which shows an extremum point of the gravitational acceleration at $\xi=1/
\sqrt{2}\approx0.707\,106\,781$, in agreement with Eqs.\,(\ref{eq:gep5}) and 
(\ref{eq:PPLE}). 

\subsection{Hernquist density profiles}
\label{hern}

Hernquist  density profiles (Hernquist 1990), hereafter quoted as H density
profiles, are defined as (e.g., Caimmi and Valentinuzzi 2008):
\begin{equation}
\label{eq:fH}
f(\xi)=\frac8{\xi(1+\xi)^3}~~;
\end{equation}
where the scaling logarithmic slope reads:
\begin{equation}
\label{eq:pH}
\left(\frac{\diff\log f}{\diff\log\xi}\right)_{\log\xi=0}=-\frac52~~;
\end{equation}
and the substitution of Eq.\,(\ref{eq:fH}) into (\ref{eq:M})-(\ref{eq:gc})
yields after some algebra:
\begin{lefteqnarray}
\label{eq:FH}
&& F(\xi)=8\left[\frac1{(1+\xi)^2}-\frac1{(1+\Xi)^2}\right]~~; \\
\label{eq:numH}
&& \nu_{\rm mas}(\xi)=\frac{12\xi^2}{(1+\xi)^2}~~; \\
\label{eq:rhomH}
&& \frac{\bar{\rho}(r)}{\rho^\dagger}=\frac{12}{\xi(1+\xi)^2}~~; \\
\label{eq:rmrH}
&& \frac{\bar{\rho}(r)}{\rho(r)}=\frac32(1+\xi)~~; \\
\label{eq:v2H}
&& \left[\frac{v(r)}{v^\dagger}\right]^2=\frac{12\xi}{(1+\xi)^2}~~; \\
\label{eq:gcsiH}
&& \frac{g(r)}{g^\dagger}=\frac{12}{(1+\xi)^2}~~;
\end{lefteqnarray}
finally, the particularization of Eq.\,(\ref{eq:gest}) to the case under
consideration reads:
\begin{equation}
\label{eq:gestH}
\frac32(1+\xi)=\frac32~~;
\end{equation}
which implies no extremum point for the gravitational acceleration inside the
domain.   In addition, Eq.\,(\ref{eq:gcsiH}) discloses that a finite
gravitational acceleration occurs even if a test particle lies infinitely
close to the centre.

\subsection{Generalized power-law density profiles}
\label{gpdp}

Generalized power-law density profiles are defined as (e.g., Bottazzi 2011):
\begin{equation}
\label{eq:fgb}
f(\xi)=\frac{C_\gamma+1}{C_\gamma+\xi^\gamma}\frac{(C_\alpha+1)^\chi}
{(C_\alpha+\xi^\alpha)^\chi}~~;
\end{equation}
where the following families are worth of note.
\begin{description}
\item[$\bullet$]
$C_\gamma=0$; $C_\alpha=0$; generalized power-law density profiles reduce to
pure power-law density profiles with exponent, $\beta$, expressed as:
\begin{equation}
\label{eq:beta}
\beta=\gamma+\alpha\chi~~;\qquad\chi=\frac{\beta-\gamma}\alpha~~;
\end{equation}
which shows the meaning of the exponent, $\chi$, appearing in
Eq.\,(\ref{eq:fgb}).
\item[$\bullet$]
$C_\gamma=0$; $C_\alpha=1$; generalized power-law density profiles reduce to
a subclass analysed in earlier attempts (Hernquist 1990; Zhao 1996) hereafter
quoted as Z density profiles.   Special cases are P density profiles,
$(\alpha,\chi,\gamma)=(2,5/2,0)$; H density profiles, $(\alpha,\chi,\gamma)=
(1,3,1)$; pseudo isothermal density profiles (e.g., Begeman et al. 1991),
$(\alpha,\chi,\gamma)=(2,1,0)$, hereafter quoted as I density profiles;
generalized pseudo isothermal density profiles (Spano et al. 2008), 
$(\alpha,\chi,\gamma)=(2,3/2,0)$, hereafter quoted as S density profiles.
\item[$\bullet$]
$C_\gamma=1$; $C_\alpha=0$; generalized power-law density profiles reduce to
a subclass of Z density profiles, $(\alpha,\chi,\gamma)=(\gamma,1,\alpha\chi)
$, including I density profiles.
\item[$\bullet$]
$C_\gamma=1$; $C_\alpha=1$; generalized power-law density profiles reduce to
a subclass of density profiles including a special case analysed in an earlier
attempt (Burkert 1995), $(\alpha,\chi,\gamma)=(2,1,1)$, hereafter
quoted as B density profiles.
\end{description}

For generalized power-law density profiles, the scaling logarithmic slope
reads:
\begin{equation}
\label{eq:pbg}
\left(\frac{\diff\log f}{\diff\log\xi}\right)_{\log\xi=0}=-\frac\gamma
{C_\gamma+1}-\frac{\beta-\gamma}{C_\alpha+1}~~;
\end{equation}
which, for Z density profiles, reduces to:
\begin{equation}
\label{eq:pZ}
\left(\frac{\diff\log f}{\diff\log\xi}\right)_{\log\xi=0}=-\frac{\beta+\gamma}
2~~;
\end{equation}
the geometrical meaning of the dimensionless scaling radius is analysed in
Appendix \ref{a:geme}.

The substitution of Eq.\,(\ref{eq:fgb}) into (\ref{eq:M})-(\ref{eq:gc})
yields, in general, integrals which cannot be analytically calculated.   For
this reason, further considerations shall be restricted to the above mentioned
special cases, where the primitive functions can be explicitly expressed.

\subsection{Pseudo isothermal density profiles}
\label{psI}

For I density profiles, Eq.\,(\ref{eq:fgb}) reduces to:
\begin{equation}
\label{eq:fI}
f(\xi)=\frac2{1+\xi^2}~~;
\end{equation}
where the scaling logarithmic slope reads:
\begin{equation}
\label{eq:pI}
\left(\frac{\diff\log f}{\diff\log\xi}\right)_{\log\xi=0}=-1~~;
\end{equation}
according to Eq.\,(\ref{eq:pZ}).

The substitution of Eq.\,(\ref{eq:fI}) into (\ref{eq:M})-(\ref{eq:gc})
yields after some algebra:
\begin{lefteqnarray}
\label{eq:FI}
&& F(\xi)=2\ln\frac{1+\Xi^2}{1+\xi^2}~~; \\
\label{eq:numI}
&& \nu_{\rm mas}(\xi)=6(\xi-\arctan\xi)~~; \\
\label{eq:rhomI}
&& \frac{\bar{\rho}(r)}{\rho^\dagger}=\frac6{\xi^3}(\xi-\arctan\xi)~~; \\
\label{eq:rmrI}
&& \frac{\bar{\rho}(r)}{\rho(r)}=\frac3{\xi^3}(1+\xi^2)(\xi-\arctan\xi)~~; \\
\label{eq:v2I}
&& \left[\frac{v(r)}{v^\dagger}\right]^2=\frac6\xi(\xi-\arctan\xi)~~; \\
\label{eq:gcsiI}
&& \frac{g(r)}{g^\dagger}=\frac6{\xi^2}(\xi-\arctan\xi)~~;
\end{lefteqnarray}
finally, the particularization of Eq.\,(\ref{eq:gest}) to the case under
consideration reads:
\begin{equation}
\label{eq:gestI}
\arctan\xi=\frac{\xi(2+\xi^2)}{2(1+\xi^2)}~~;
\end{equation}
and the extremum point for the gravitational acceleration is defined by the
solution of Eq.\,(\ref{eq:gestI}), as $\xi\approx1.514\,994\,606$.

\subsection{Spano et al. density profiles}
\label{psS}

For S density profiles, Eq.\,(\ref{eq:fgb}) reduces to:
\begin{equation}
\label{eq:fS}
f(\xi)=\frac{2^{3/2}}{(1+\xi^2)^{3/2}}~~;
\end{equation}
where the scaling logarithmic slope reads:
\begin{equation}
\label{eq:pS}
\left(\frac{\diff\log f}{\diff\log\xi}\right)_{\log\xi=0}=-\frac32~~;
\end{equation}
according to Eq.\,(\ref{eq:pZ}).

The substitution of Eq.\,(\ref{eq:fS}) into (\ref{eq:M})-(\ref{eq:gc})
yields after some algebra:
\begin{lefteqnarray}
\label{eq:FS}
&& F(\xi)=2^{5/2}\left(\frac1{\sqrt{1+\xi^2}}-\frac1{\sqrt{1+\Xi^2}}\right)~~;
\\
\label{eq:numS}
&& \nu_{\rm mas}(\xi)=6\sqrt{2}\left(\arcsinh\xi-\frac\xi{\sqrt{1+\xi^2}}
\right)~~; \\
\label{eq:rhomS}
&& \frac{\bar{\rho}(r)}{\rho^\dagger}=\frac{6\sqrt{2}}{\xi^3}\left(\arcsinh\xi
-\frac\xi{\sqrt{1+\xi^2}}\right)~~; \\
\label{eq:rmrS}
&& \frac{\bar{\rho}(r)}{\rho(r)}=\frac{3\sqrt{2}}{\xi^3}(1+\xi^2)^{3/2}\left(
\arcsinh\xi-\frac\xi{\sqrt{1+\xi^2}}\right)~~; \\
\label{eq:v2S}
&& \left[\frac{v(r)}{v^\dagger}\right]^2=\frac{6\sqrt{2}}\xi\left(\arcsinh\xi-
\frac\xi{\sqrt{1+\xi^2}}\right)~~; \\
\label{eq:gcsiS}
&& \frac{g(r)}{g^\dagger}=\frac{6\sqrt{2}}{\xi^2}\left(\arcsinh\xi-\frac\xi
{\sqrt{1+\xi^2}}\right)~~;
\end{lefteqnarray}
finally, the particularization of Eq.\,(\ref{eq:gest}) to the case under
consideration reads:
\begin{equation}
\label{eq:gestS}
\arcsinh\xi=\frac{\xi(2+3\xi^2)}{2(1+\xi^2)^{3/2}}~~;
\end{equation}
and the extremum point for the gravitational acceleration is defined by the
solution of Eq.\,(\ref{eq:gestS}), as $\xi\approx1.027\,115\,657$. 

\subsection{Burkert density profiles}
\label{psB}

For B density profiles, Eq.\,(\ref{eq:fgb}) reduces to:
\begin{equation}
\label{eq:fB}
f(\xi)=\frac2{1+\xi}\frac2{1+\xi^2}~~;
\end{equation}
where the scaling logarithmic slope reads:
\begin{equation}
\label{eq:pB}
\left(\frac{\diff\log f}{\diff\log\xi}\right)_{\log\xi=0}=-\frac32~~;
\end{equation}
according to Eq.\,(\ref{eq:pZ}).

The substitution of Eq.\,(\ref{eq:fB}) into (\ref{eq:M})-(\ref{eq:gc})
yields after some algebra:
\begin{lefteqnarray}
\label{eq:FB}
&& F(\xi)=-2\ln\frac{1+\xi^2}{1+\Xi^2}+4\ln\frac{1+\xi}{1+\Xi}+4(\arctan\Xi-
\arctan\xi)~~; \\
\label{eq:numB}
&& \nu_{\rm mas}(\xi)=3[\ln(1+\xi^2)+2\ln(1+\xi)-2\arctan\xi]~~; \\
\label{eq:rhomB}
&& \frac{\bar{\rho}(r)}{\rho^\dagger}=\frac3{\xi^3}[\ln(1+\xi^2)+2\ln(1+\xi)-
2\arctan\xi]~~; \\
\label{eq:rmrB}
&& \frac{\bar{\rho}(r)}{\rho(r)}=\frac34\frac1{\xi^3}[\ln(1+\xi^2)+2\ln(1+\xi)
-2\arctan\xi]~~; \\
\label{eq:v2B}
&& \left[\frac{v(r)}{v^\dagger}\right]^2=\frac3\xi[\ln(1+\xi^2)+2\ln(1+\xi)-2
\arctan\xi]~~; \\
\label{eq:gcsiB}
&& \frac{g(r)}{g^\dagger}=\frac3{\xi^2}[\ln(1+\xi^2)+2\ln(1+\xi)-2\arctan\xi]
~~;
\end{lefteqnarray}
finally, the particularization of Eq.\,(\ref{eq:gest}) to the case under
consideration reads:
\begin{equation}
\label{eq:gestB}
\ln(1+\xi^2)+2\ln(1+\xi)-2\arctan\xi=\frac{2\xi^3}{(1+\xi)(1+\xi^2)}~~;
\end{equation}
and the extremum point for the gravitational acceleration is defined by the
solution of Eq.\,(\ref{eq:gestB}), as $\xi\approx0.963\,398\,283$. 

\subsection{Gravitational acceleration vs. density profile}
\label{gadp}

The dependence of the gravitational acceleration on the density profile can be
related to the dimensionless effective radius, $\xi_{\rm eff}$, where
the maximum absolute value is attained.   For the density profiles considered
in the
current Section, the results are listed in Table \ref{t:gadp} together with
values of related parameters.
\begin{table}
\caption{Parameters of one-component density profiles considered in the text:
polytropic index, $n$, for polytropes; constants, $C_\gamma$, $C_\alpha$, and
exponents, $\alpha$, $\chi$, $\gamma$, $\beta$, for generalized power-law
density profiles, Eqs.\,(\ref{eq:fgb})-(\ref{eq:beta}); and dimensionless
effective radius, $\xi_{\rm eff}$.   An extremum point for the
gravitational acceleration occurs only if $0<\xi_{\rm eff}<\Xi$. For polytropic
density profiles, $\xi_{\rm eff}=\Xi$, $n=0$, and $\xi_{\rm eff}=\sqrt{3/2}$,
$n=5$, respectively.   For Plummer density profile, $\xi_{\rm eff}=1/\sqrt2$.
The notations specifying power-law density profiles, (0-2) and (2-3), are to
be intended as $(0\le b\le2)$ and $(2\le b\le3)$, respectively.   See text for
further details.}
\label{t:gadp}
\begin{center}
\begin{tabular}{lllllllll} \hline
\multicolumn{1}{c}{density profile}
&\multicolumn{1}{c}{$n$}
&\multicolumn{1}{c}{$C_\gamma$}
&\multicolumn{1}{c}{$C_\alpha$}
&\multicolumn{1}{c}{$\alpha$}
&\multicolumn{1}{c}{$\chi$} 
&\multicolumn{1}{c}{$\gamma$}
&\multicolumn{1}{c}{$\beta$}
& \multicolumn{1}{c}{$\xi_{\rm eff}$} \\
\noalign{\smallskip}
pure  power-law (0-2) &   & 0 & 0 & $\alpha$ & $\frac{\beta-\gamma}{\alpha}$ & $\gamma$ & $\beta$ &  $0$                                                              \\
pure  power-law (2-3) &   & 0 & 0 & $\alpha$ & $\frac{\beta-\gamma}{\alpha}$ & $\gamma$ & $\beta$ &  $\Xi$                                                            \\
cored power-law (0-2) &   &   &   &          &                         &          &         &  $\min\left[\left(\frac23\frac{3-b}{2-b}\right)^{1/b},\Xi\right]$ \\
cored power-law (2-3) &   &   &   &          &                         &          &         &  $\Xi$                                                            \\
polytropic            & 0 &   &   &          &                         &          &         &  $2.449\,489\,743$                                                \\
polytropic            & 1 &   &   &          &                         &          &         &  $2.081\,575\,99$                                                 \\
polytropic            & 5 &   &   &          &                         &          &         &  $1.224\,744\,871$                                                \\
Plummer (1911)        &   & 0 & 1 & 2        & 2.5                     & 0        & 5       &  $0.707\,106\,781$                                                \\
Hernquist (1990)      &   & 0 & 1 & 1        & 3.0                     & 1        & 4       &  $0$                                                              \\
Begeman et al. (1991)  &   & 0 & 1 & 2        & 1.0                     & 0        & 2       &  $1.514\,994\,606$                                                \\
Spano et al. (2008)   &   & 0 & 1 & 2        & 1.5                     & 0        & 3       &  $1.027\,115\,657$                                                \\
Burkert (1995)        &   & 1 & 1 & 2        & 1.0                     & 1        & 3       &  $0.963\,398\,283$                                                \\
\noalign{\smallskip}      
\hline                                                       
\end{tabular}                                                
\end{center}                                                 
\end{table}                                                  
An inspection of Table \ref{t:gadp} shows the occurrence of an extremum point
(maximum absolute value) in the gravitational acceleration distribution, with the exception
of pure power-law, cored power-law $(2\le b\le3)$, H density profiles, and the
possible exception of cored power-law $(0\le b<2)$ density profiles, according
to the selected dimensionless truncation radius, $\Xi$.

\section{Two-component density  profiles}
\label{2cdp}

Let $\rho_i(r)$, $\rho_j(r)$, be generic but concentric homeoidally striated
spherical density profiles, and $M_i(r)$, $M_j(r)$, related masses enclosed
within the distance, $r$, from the centre.   The total
mass, the squared circular velocity, and the gravitational acceleration are
expressed by Eqs.\,(\ref{eq:Mr})-(\ref{eq:g}) where $\rho(r)=\rho_i(r)+\rho_j
(r)$, $M(r)=M_i(r)+M_j(r)$, which implies $\bar{\rho}(r)=\bar{\rho}_i(r)+
\bar{\rho}_j(r)$.

It can be seen the validity of Eq.\,(\ref{eq:rhor}) for
each subsystem i.e. the existence of an extremum point of related
gravitational acceleration, $g_u=-GM_u(r)/r^2$, $u=i,j$, is a necessary
condition for the validity of Eq.\,(\ref{eq:rhor}) for the whole system i.e.
the existence of an extremum point of related gravitational acceleration,
$g=g_i+g_j$.

In dimensionless coordinates, generic density profiles are expressed by
Eqs.\,(\ref{eq:rho})-(\ref{eq:gest}), provided each quantity is indexed by
$i$ or $j$, according to the selected subsystem.  In addition, the following
relations hold (e.g., Caimmi and Valentinuzzi 2008):
\begin{leftsubeqnarray}
\slabel{eq:qija}
&& \xi_i=y^\dagger\xi_j~~;\qquad\frac{\Xi_j}{\Xi_i}=\frac y{y^\dagger}~~;
\qquad\frac{\nu_{j,{\rm mas}}(r)}{\nu_{i,{\rm mas}}(r)}=\frac{m(r)}{m^\dagger}
~~; \\
\slabel{eq:qijb}
&& y=\frac{R_j}{R_i}~~;\qquad y^\dagger=\frac{r_j^\dagger}{r_i^\dagger}~~;
\qquad m(r)=\frac{M_j(r)}{M_i(r)}~~;\qquad m^\dagger=\frac{M_j^\dagger}
{M_i^\dagger}~~;\qquad \\
\slabel{eq:qijc}
&& \frac{\rho_j^\dagger}{\rho_i^\dagger}=\frac{m^\dagger}{(y^\dagger)^3}~~;
\label{seq:qij}
\end{leftsubeqnarray}
where $y\ge1$ without loss of generality, which implies $i$ is the inner
subsystem and $j$ the outer unless their surfaces coincide.

Using Eqs.\,(\ref{eq:M})-(\ref{eq:gc}) and (\ref{seq:qij}), the mass enclosed
within the radius, $r$, and related squared circular velocity and
gravitational acceleration are expressed as:
\begin{lefteqnarray}
\label{eq:M2}
&& M(r)=M_i^\dagger\nu_{i,{\rm mas}}(\xi_i)+M_j^\dagger\nu_{j,{\rm mas}}
(\xi_j)~~; \\
\label{eq:v22}
&& v^2(r)=(v_i^\dagger)^2\frac{\nu_{i,{\rm mas}}(\xi_i)}{\xi_i}+
(v_j^\dagger)^2\frac{\nu_{j,{\rm mas}}(\xi_j)}{\xi_j}~~; \\
\label{eq:gcs2}
&& g(r)=g_i^\dagger\frac{\nu_{i,{\rm mas}}(\xi_i)}{\xi_i^2}+
g_j^\dagger\frac{\nu_{j,{\rm mas}}(\xi_j)}{\xi_j^2}~~;
\end{lefteqnarray}
and the extremum point of the gravitational acceleration, via
Eqs.\,(\ref{eq:rhor})-(\ref{eq:gest}) and (\ref{seq:qij}), is the solution 
of the following equation:
\begin{equation}
\label{eq:ges2}
\rho_i^\dagger\left[\frac{\nu_{i,{\rm mas}}(\xi_i)}{\xi_i^3}-\frac32f_i(\xi_i)
\right]+\rho_j^\dagger\left[\frac{\nu_{j,{\rm mas}}(\xi_j)}{\xi_j^3}-\frac32
f_j(\xi_j)\right]=0~~;
\end{equation}
where $\xi_i$ and $\xi_j$ are related via
Eqs.\,(\ref{eq:qija})-(\ref{eq:qijb}).

Some special density distributions shall be considered in the following, and
Eq.\,(\ref{eq:ges2}) shall be written in terms of the additional parameters,
$m^\dagger$ and $y^\dagger$, from which the extremum points of the
gravitational acceleration also depend.

\subsection{PI density profiles}
\label{PI}

In the special case of an inner P density profile, $i=$P, and an outer I
density profile, $j=$I, the substitution of
Eqs.\,(\ref{eq:fP})-(\ref{eq:gcsiP}) and (\ref{eq:fI})-(\ref{eq:gcsiI}) into
(\ref{seq:qij})-(\ref{eq:gcs2}) yields:
\begin{lefteqnarray}
\label{eq:cPI}
&& \xi_{\rm P}=y^\dagger\xi_{\rm I}~~; \\
\label{eq:mPI}
&& m(r)=m^\dagger\frac{6(\xi_{\rm I}-\arctan\xi_{\rm I})}{2^{5/2}\xi_{\rm P}^
3(1+\xi_{\rm P}^2)^{-3/2}}~~; \\
\label{eq:MPI}
&& M(r)=M_{\rm P}^\dagger\frac{2^{5/2}\xi_{\rm P}^3}{(1+\xi_{\rm P}^2)^{3/2}}+
M_{\rm I}^\dagger6(\xi_{\rm I}-\arctan\xi_{\rm I})~~; \\
\label{eq:v2PI}
&& v^2(r)=(v_{\rm P}^\dagger)^2\frac{2^{5/2}\xi_{\rm P}^2}{(1+\xi_{\rm P}^2)^
{3/2}}+(v_{\rm I}^\dagger)^2\frac{6(\xi_{\rm I}-\arctan\xi_{\rm I})}{\xi_
{\rm I}}~~; \\
\label{eq:gcsPI}
&& g(r)=g^\dagger_{\rm P}\frac{2^{5/2}\xi_{\rm P}}{(1+\xi_{\rm P}^2)^{3/2}}+
g^\dagger_{\rm I}\frac{6(\xi_{\rm I}-\arctan\xi_{\rm I})}{\xi_{\rm I}^2}~~;
\end{lefteqnarray}
and the extremum point of the gravitational acceleration is the solution of
Eq.\,(\ref{eq:ges2}) particularized to the case under consideration, which
after some algebra reads:
\begin{equation}
\label{eq:gePI}
\frac{2^{3/2}\rho_{\rm P}^\dagger}{(1+\xi_{\rm P}^2)^{3/2}}\frac{2\xi_{\rm P}^
2-1}{1+\xi_{\rm P}^2}+3\rho_{\rm I}^\dagger\left[\frac{2+\xi_{\rm I}^2}{\xi_
{\rm I}^2(1+\xi_{\rm I}^2)}-\frac{2\arctan\xi_{\rm I}}{\xi_{\rm I}^3}\right]=0
~~;
\end{equation}
that shows an additional dependence on the parameters, $m^\dagger$,
$y^\dagger$, via Eqs.\,(\ref{eq:qijc}) and (\ref{eq:cPI}), respectively.

\subsection{HI density profiles}
\label{HI}

In the special case of an inner H density profile, $i=$H, and an outer I
density profile, $j=$I, the substitution of
Eqs.\,(\ref{eq:fH})-(\ref{eq:gcsiH}) and (\ref{eq:fI})-(\ref{eq:gcsiI}) into
(\ref{seq:qij})-(\ref{eq:gcs2}) yields:
\begin{lefteqnarray}
\label{eq:cHI}
&& \xi_{\rm H}=y^\dagger\xi_{\rm I}~~; \\
\label{eq:mHI}
&& m(r)=m^\dagger\frac{\xi_{\rm I}-\arctan\xi_{\rm I}}{2\xi_{\rm H}^
2(1+\xi_{\rm H})^{-2}}~~; \\
\label{eq:MHI}
&& M(r)=M_{\rm H}^\dagger\frac{12\xi_{\rm H}^2}{(1+\xi_{\rm H})^2}+
M_{\rm I}^\dagger6(\xi_{\rm I}-\arctan\xi_{\rm I})~~; \\
\label{eq:v2HI}
&& v^2(r)=(v_{\rm H}^\dagger)^2\frac{12\xi_{\rm H}}{(1+\xi_{\rm H})^2}+
(v_{\rm I}^\dagger)^2\frac{6(\xi_{\rm I}-\arctan\xi_{\rm I})}{\xi_{\rm I}}~~; \\
\label{eq:gcsHI}
&& g(r)=g^\dagger_{\rm H}\frac{12}{(1+\xi_{\rm H})^2}+
g^\dagger_{\rm I}\frac{6(\xi_{\rm I}-\arctan\xi_{\rm I})}{\xi_{\rm I}^2}~~;
\end{lefteqnarray}
and the extremum point of the gravitational acceleration is the solution of
Eq.\,(\ref{eq:ges2}) particularized to the case under consideration, which
after some algebra reads:
\begin{equation}
\label{eq:geHI}
\frac{12\rho_{\rm H}^\dagger}{(1+\xi_{\rm H})^3}+3\rho_{\rm I}^\dagger\left[
\frac{2+\xi_{\rm I}^2}{\xi_{\rm I}^2(1+\xi_{\rm I}^2)}-\frac{2\arctan
\xi_{\rm I}}{\xi_{\rm I}^3}\right]=0~~;
\end{equation}
that shows an additional dependence on the parameters, $m^\dagger$,
$y^\dagger$, via Eqs.\,(\ref{eq:qijc}) and (\ref{eq:cHI}), respectively.

\subsection{PS density profiles}
\label{PS}

In the special case of an inner P density profile, $i=$P, and an outer S
density profile, $j=$S, the substitution of
Eqs.\,(\ref{eq:fP})-(\ref{eq:gcsiP}) and (\ref{eq:fS})-(\ref{eq:gcsiS}) into
(\ref{seq:qij})-(\ref{eq:gcs2}) yields:
\begin{lefteqnarray}
\label{eq:cPS}
&& \xi_{\rm P}=y^\dagger\xi_{\rm S}~~; \\
\label{eq:mPS}
&& m(r)=m^\dagger\frac{6\sqrt{2}[\arcsinh\xi_{\rm S}-\xi_{\rm S}(1+
\xi_{\rm S}^2)^{-1/2}]}{2^{5/2}\xi_{\rm P}^3(1+\xi_{\rm P}^2)^{-3/2}}~~; \\
\label{eq:MPS}
&& M(r)=M_{\rm P}^\dagger\frac{2^{5/2}\xi_{\rm P}^3}{(1+\xi_{\rm P}^2)^{3/2}}+
M_{\rm S}^\dagger6\sqrt{2}\left(\arcsinh\xi_{\rm S}-\frac{\xi_{\rm S}}
{\sqrt{1+\xi_{\rm S}^2}}\right)~~; \\
\label{eq:v2PS}
&& v^2(r)=(v_{\rm P}^\dagger)^2\frac{2^{5/2}\xi_{\rm P}^2}{(1+\xi_{\rm P}^2)^
{3/2}}+(v_{\rm S}^\dagger)^26\sqrt{2}\left(\frac{\arcsinh\xi_{\rm S}}
{\xi_{\rm S}}-\frac1{\sqrt{1+\xi_{\rm S}^2}}\right)~~; \\
\label{eq:gcsPS}
&& g(r)=g^\dagger_{\rm P}\frac{2^{5/2}\xi_{\rm P}}{(1+\xi_{\rm P}^2)^{3/2}}+
g^\dagger_{\rm S}\frac{6\sqrt{2}}{\xi_{\rm S}}\left(\frac{\arcsinh\xi_{\rm S}}
{\xi_{\rm S}}-\frac1{\sqrt{1+\xi_{\rm S}^2}}\right)~~;
\end{lefteqnarray}
and the extremum point of the gravitational acceleration is the solution of
Eq.\,(\ref{eq:ges2}) particularized to the case under consideration, which
after some algebra reads:
\begin{equation}
\label{eq:gePS}
\frac{2^{3/2}\rho_{\rm P}^\dagger}{(1+\xi_{\rm P}^2)^{3/2}}\frac{2\xi_{\rm P}^
2-1}{1+\xi_{\rm P}^2}+\frac{3\sqrt{2}\rho_{\rm S}^\dagger}{\xi_{\rm S}^2}
\left[2\frac{\arcsinh\xi_{\rm S}}{\xi_{\rm S}}-\frac{2+3\xi_{\rm S}^2}
{(1+\xi_{\rm S}^2)^{3/2}}\right]=0~~;
\end{equation}
that shows an additional dependence on the parameters, $m^\dagger$,
$y^\dagger$, via Eqs.\,(\ref{eq:qijc}) and (\ref{eq:cPS}), respectively.

\subsection{HS density profiles}
\label{HS}

In the special case of an inner H density profile, $i=$H, and an outer S
density profile, $j=$S, the substitution of
Eqs.\,(\ref{eq:fH})-(\ref{eq:gcsiH}) and (\ref{eq:fS})-(\ref{eq:gcsiS}) into
(\ref{seq:qij})-(\ref{eq:gcs2}) yields:
\begin{lefteqnarray}
\label{eq:cHS}
&& \xi_{\rm H}=y^\dagger\xi_{\rm S}~~; \\
\label{eq:mHS}
&& m(r)=m^\dagger\frac{6\sqrt{2}[\arcsinh\xi_{\rm S}-\xi_{\rm S}(1+\xi_{\rm S}
^2)^{-1/2}]}{12\xi_{\rm H}^2(1+\xi_{\rm H})^{-2}}~~; \\
\label{eq:MHS}
&& M(r)=M_{\rm H}^\dagger\frac{12\xi_{\rm H}^2}{(1+\xi_{\rm H})^2}+
M_{\rm S}^\dagger6\sqrt{2}\left(\arcsinh\xi_{\rm S}-\frac{\xi_{\rm S}}
{\sqrt{1+\xi_{\rm S}^2}}\right)~~; \\
\label{eq:v2HS}
&& v^2(r)=(v_{\rm H}^\dagger)^2\frac{12\xi_{\rm H}}{(1+\xi_{\rm H})^2}+
(v_{\rm S}^\dagger)^26\sqrt{2}\left(\frac{\arcsinh\xi_{\rm S}}
{\xi_{\rm S}}-\frac1{\sqrt{1+\xi_{\rm S}^2}}\right)~~; \\
\label{eq:gcsHS}
&& g(r)=g^\dagger_{\rm H}\frac{12}{(1+\xi_{\rm H})^2}+
g^\dagger_{\rm S}\frac{6\sqrt{2}}{\xi_{\rm S}}\left(\frac{\arcsinh\xi_{\rm S}}
{\xi_{\rm S}}-\frac1{\sqrt{1+\xi_{\rm S}^2}}\right)~~;
\end{lefteqnarray}
and the extremum point of the gravitational acceleration is the solution of
Eq.\,(\ref{eq:ges2}) particularized to the case under consideration, which
after some algebra reads:
\begin{equation}
\label{eq:geHS}
\frac{12\rho_{\rm H}^\dagger}{(1+\xi_{\rm H})^3}+\frac{3\sqrt{2}\rho_{\rm S}^
\dagger}{\xi_{\rm S}^2}\left[2\frac{\arcsinh\xi_{\rm S}}{\xi_{\rm S}}-\frac
{2+3\xi_{\rm S}^2}{(1+\xi_{\rm S}^2)^{3/2}}\right]=0~~;
\end{equation}
that shows an additional dependence on the parameters, $m^\dagger$,
$y^\dagger$, via Eqs.\,(\ref{eq:qijc}) and (\ref{eq:cHS}), respectively.

\section{Tidal effects on embedded subsystems}
\label{app}

Let a subsystem be completely embedded within another one.   Let the former
and the latter be hereafter quoted as the embedded and the embedding
sphere, respectively, owing to the assumption of spherical symmetry.
Accordingly, related centres of mass shall be hereafter quoted as centres,
keeping in mind the centre of mass coincides with the geometrical centre in
the case under consideration.  The
dependence of the tidal radius of the embedded sphere on the density profile
of the embedding sphere shall be exploited below for both one-component and
two-component systems, having in mind globular clusters within galaxies and
dark matter haloes.

To this aim, the tidal radius must be clearly defined and
considerations be restricted to density profiles (among those listed in Table
\ref{t:gadp}) which satisfactorily fit to observed or inferred matter
distributions in galaxies and/or dark matter haloes, namely P, H, I, S, B, for
one-component systems while, on the other
hand, PI, HI, PS, HS, are acceptable for two-component systems.

\subsection{Tidal radius}
\label{tira}

Let the embedded and the embedding sphere be conceived as a globular cluster
and a galaxy, respectively, both assumed spherical-symmetric.   The tidal
radius of the embedded sphere can be defined by use of an either local (e.g.,
von Hoerner 1958; Vesperini 1997; Brosche et al. 1999; Gajda and Lokas 2015)
or global (e.g., Caimmi
and Secco 2003) criterion, where the former shall be preferred here in that it
involves a single test particle instead of the embedded sphere as a whole.

\begin{figure*}[t]
\begin{center}
\includegraphics[scale=0.8]{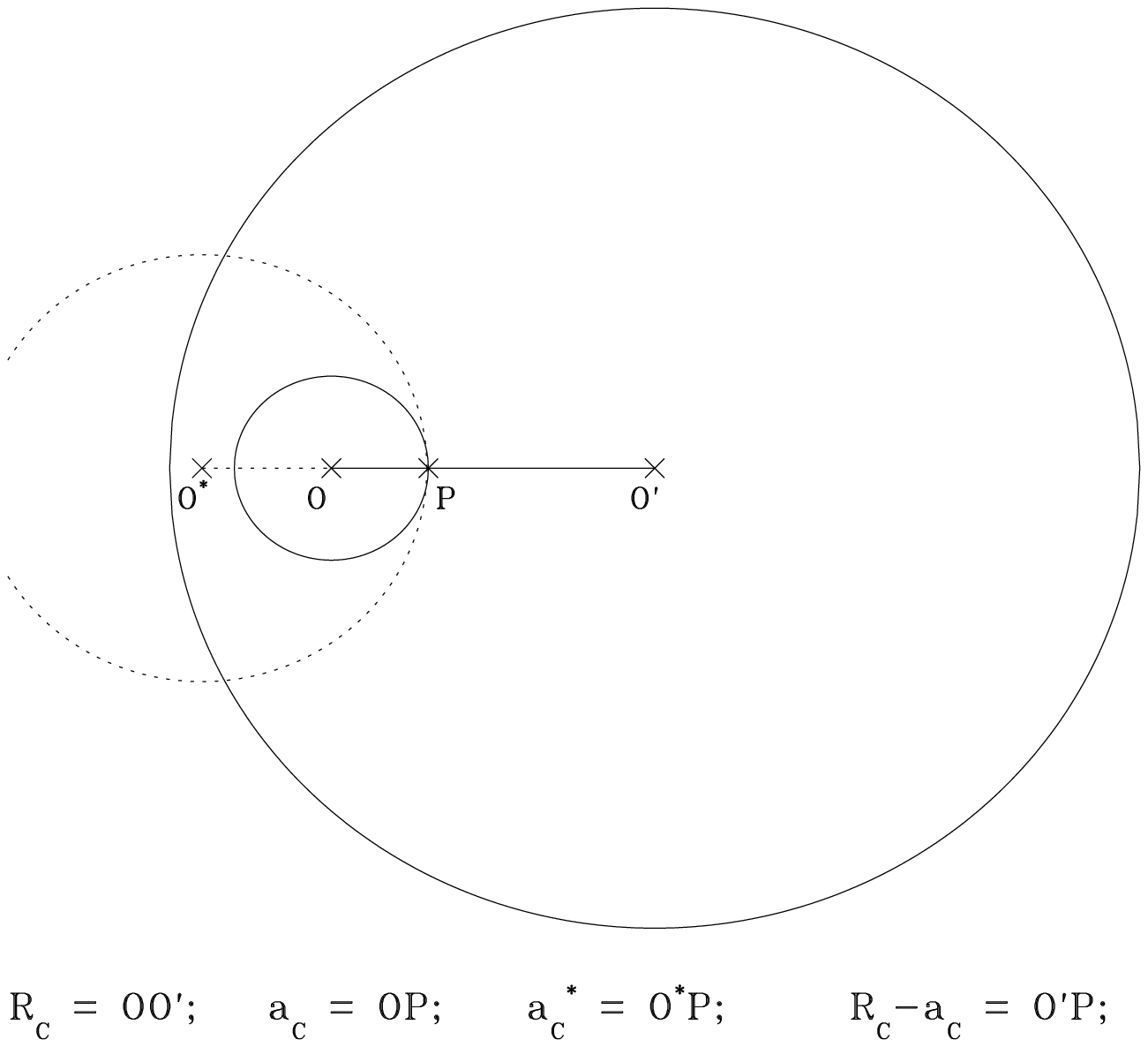}
\caption[OKND]{The embedding and the embedded sphere, centred in
${\sf O}^\prime$ and ${\sf O}$, respectively.   The minimum gravitational
force exerted on the surface of the embedded sphere takes place on
${\sf P}$, where the attraction due to each subsystem acts along the
same direction but on opposite sides.   A null gravitational force on
${\sf P}$ defines $a_{\rm C}$ as the tidal radius of the embedded
sphere, $a_{\rm C}^\ast$, where ${\sf O}$ moves to ${\sf O}^\ast$
while ${\sf P}$ remains unchanged.  Points mentioned above are represented as
saltires.   The figure is not in scale for sake of clarity: the embedded
sphere, even if the radius equals the tidal radius (dotted), is completely
lying within the embedding sphere.   See text for further details.}
\label{f:tidi}
\end{center}
\end{figure*}

Let $M_{\rm G}(R)$, $a_{\rm G}$, be the mass profile and the truncation
radius, respectively, of the embedding sphere, and
$M_{\rm C}$, $a_{\rm C}$, the mass and the truncation radius, respectively, of
the embedded sphere.   Let ${\sf P}$ be the intersection point between
the surface of the embedded sphere and the segment,
$\overline{{\sf OO}^\prime}$, joining the centre of the
embedded and the embedding sphere, respectively.   A test particle of unit
mass placed on ${\sf P}$ is subjected to a gravitational force:
\begin{equation}
\label{eq:FGP}
F_{\rm C}({\sf P})=\frac{GM_{\rm C}}{a_{\rm C}^2}~~;
\end{equation}
due to the embedded sphere, and:
\begin{equation}
\label{eq:FCP}
F_{\rm G}({\sf P})=-\frac{GM_{\rm G}(R_{\rm C}-a_{\rm C})}{(R_{\rm C}-
a_{\rm C})^2}~~;
\end{equation}
due to the embedding sphere, where $R_{\rm C}$,
$a_{\rm C}\le R_{\rm C}\le a_{\rm G}-a_{\rm C}$ is the distance
between the centre of the embedded and the embedding sphere,
respectively, as depicted in Fig.\,\ref{f:tidi}.

Let the tidal radius of the embedded sphere, $a_{\rm C}^\ast$,
be defined as related to a null resulting gravitational attraction, 
$F_{\rm G}({\sf P})+F_{\rm C}({\sf P})=0$, or more explicitly:
\begin{equation}
\label{eq:greq}
\frac{GM_{\rm C}}{a_{\rm C}^2}=
\frac{GM_{\rm G}(R_{\rm C}-a_{\rm C})}{(R_{\rm C}-a_{\rm C})^2}~~;
\end{equation}
which, using Eqs.\,(\ref{eq:M}) and (\ref{eq:Mc}), is equivalent to:
\begin{lefteqnarray}
\label{eq:ganu}
&& \left(\frac{1-\gamma}\gamma\right)^2=\frac{\nu_{\rm mas}(\xi_{\rm C}-
\delta_{\rm C})}{(\mu^\dagger)^2}~~; \\
\label{eq:gamu}
&& \xi_{\rm C}=\frac{R_{\rm C}}{r_{\rm G}^\dagger}~~;\quad
\delta_{\rm C}=\frac{a_{\rm C}}{r_{\rm G}^\dagger}~~;\quad
\gamma=\frac{a_{\rm C}}{R_{\rm C}}=\frac{\delta_{\rm C}}{\xi_{\rm C}}~~;\quad
\mu^\dagger=\left(\frac{M_{\rm C}}{M_{\rm G}^\dagger}\right)^{1/2}~~;\qquad
\end{lefteqnarray}
where the validity of Eq.\,(\ref{eq:ganu}) implies $a_{\rm C}=a_{\rm C}^\ast$
or $\gamma=\gamma^\ast$.

With regard to the embedding sphere, a nonmonotonic trend of the gravitational
acceleration, characterized by a maximum absolute value within the truncation
radius, is a
necessary condition for the occurrence of two different values of the tidal
radius of the embedded sphere: one, sufficiently close to the centre of the
embedding sphere, and one other, sufficiently close to the truncation radius.
Then the embedded sphere can be considered bound for $1\ge\gamma>
(\gamma^\ast)^-$, $0\le\gamma<(\gamma^\ast)^+$, and (partially) unbound for
$(\gamma^\ast)^+\le\gamma\le(\gamma^\ast)^-$, where the apexes, $-$ and +,
relate to the lower and higher value of the galactocentric distance,
respectively.

An instantaneous tidal radius can be defined for the embedded sphere for fixed
mass, $M_{\rm C}$, and distance, $R_{\rm C}-a_{\rm C}$, or
$\xi_{\rm C}-\delta_{\rm C}$ in dimensionless coordinates.   Accordingly,
$\gamma$
varies via $a_{\rm C}$ instead of $R_{\rm C}$ as in the former alternative,
see also Fig.\,\ref{f:tidi}.   The instantaneous tidal radius,
$\eta(\xi_{\rm C}-\delta_{\rm C})$, can be inferred from Eq.\,(\ref{eq:ganu})
as solution of a second degree equation in $\gamma$.   The result is:
\begin{lefteqnarray}
\label{eq:istr}
&& \eta^\mp(\xi_{\rm C}-\delta_{\rm C})=\frac{\mu(\xi_{\rm C}-\delta_
{\rm C})}{\mu(\xi_{\rm C}-\delta_{\rm C})\mp1}~~; \\
\label{eq:mutr}
&& \mu(\xi_{\rm C}-\delta_{\rm C})=\left[\frac{M_{\rm C}}
{M_{\rm G}(\xi_{\rm C}-\delta_{\rm C})}\right]^{1/2}=\left[\frac
{(\mu^\dagger)^2}{\nu_{\rm mas}(\xi_{\rm C}-\delta_{\rm C})}\right]^{1/2}~~;
\end{lefteqnarray}
where $\eta^-(\xi_{\rm C}-\delta_{\rm C})=\gamma_-^\ast(\xi_{\rm C}-
\delta_{\rm C})>1$, which
has no physical meaning, as shown below.   Then the acceptable solution is
$\eta^+(\xi_{\rm C}-\delta_{\rm C})=\gamma_+^\ast(\xi_{\rm C}-\delta_{\rm C})<
1$ which, for sake of simplicity, shall be denoted as $\eta$ in the following.
By definition, $\gamma/\eta=a_{\rm C}/a_{\rm C}^\ast$.

The above results can be extended to two-component embedding spheres by use of
Eq.\,(\ref{eq:M2}), keeping in mind $M_{\rm G}(R_{\rm C}-a_{\rm C})=
M_{i,{\rm G}}(R_{\rm C}-a_{\rm C})+M_{j,{\rm G}}(R_{\rm C}-a_{\rm C})$.
Accordingly, the combination of Eqs.\,(\ref{eq:M2}) and (\ref{eq:greq})
yields:
\begin{lefteqnarray}
\label{eq:gan2}
&& \left(\frac{1-\gamma}\gamma\right)^2=\frac{\nu_{i,{\rm mas}}
(\xi_{i,{\rm C}}-\delta_{i,{\rm C}})}{(\mu_i^\dagger)^2}+\frac
{\nu_{j,{\rm mas}}(\xi_{j,{\rm C}}-\delta_{j,{\rm C}})}{(\mu_j^\dagger)^2}~~; \\
\label{eq:gam2}
&& \xi_{u,{\rm C}}=\frac{R_{\rm C}}{r_{u,{\rm G}}^\dagger}~~;\quad
\delta_{u,{\rm C}}=\frac{a_{\rm C}}{r_{u,{\rm G}}^\dagger}~~;\quad
\gamma=\frac{a_{\rm C}}{R_{\rm C}}=\frac{\delta_{u,{\rm C}}}{\xi_{u,{\rm C}}}
~~;\quad\mu_u^\dagger=\left(\frac{M_{\rm C}}{M_{u,{\rm G}}^\dagger}\right)^
{1/2}~~; \nonumber \\
&& u=i,j~;
\end{lefteqnarray}
in addition, the combination of Eqs.\,(\ref{eq:M2}) and (\ref{eq:istr})
produces:
\begin{lefteqnarray}
\label{eq:ist2}
&& \eta^\mp(\xi_{u,{\rm C}}-\delta_{u,{\rm C}})=\frac{\mu(\xi_{u,{\rm C}}-
\delta_{u,{\rm C}})}{\mu(\xi_{u,{\rm C}}-\delta_{u,{\rm C}})\mp1}~~; \\
\label{eq:mut2}
&& \mu(\xi_{u,{\rm C}}-\delta_{u,{\rm C}})=\left[\frac{M_{\rm C}}
{M_{i,{\rm G}}(\xi_{i,{\rm C}}-\delta_{i,{\rm C}})}+\frac{M_{\rm C}}
{M_{j,{\rm G}}(\xi_{j,{\rm C}}-\delta_{j,{\rm C}})}\right]^{1/2} \nonumber \\
&& \phantom{\mu(\xi_{u,{\rm C}}-\delta_{u,{\rm C}})}=
\left[\frac{(\mu_i^\dagger)^2}
{\nu_{i,{\rm mas}}(\xi_{i,{\rm C}}-\delta_{i,{\rm C}})}+
\frac{(\mu_j^\dagger)^2}
{\nu_{j,{\rm mas}}(\xi_{j,{\rm C}}-\delta_{j,{\rm C}})}\right]^{1/2}~~;\qquad
\end{lefteqnarray}
where the acceptable solution, for reasons mentioned above, is
$\eta^+(\xi_{u,{\rm C}}-\delta_{u,{\rm C}})=\gamma_+^\ast(\xi_{u,{\rm C}}-
\delta_{u,{\rm C}})<1$
which, for sake of simplicity, shall be denoted as $\eta$ in the following.

\begin{figure*}[t]
\begin{center}
\includegraphics[scale=0.8]{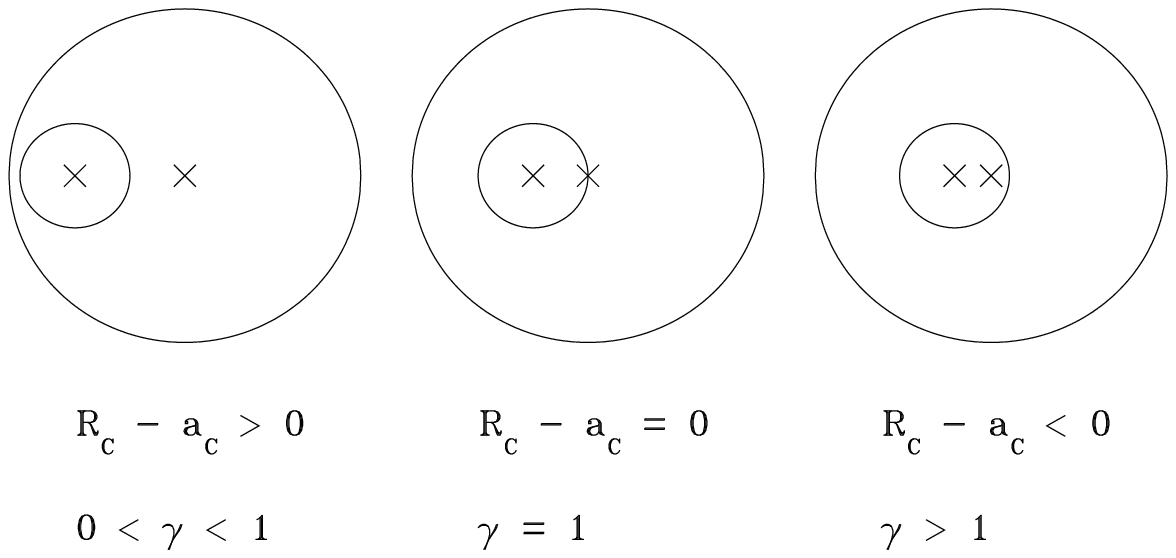}
\caption[OKND]{Mutual positions of the embedding and the embedded sphere for
different values of the inverse fractional distance,
$\gamma=a_{\rm C}/R_{\rm C}$.   Related centres are represented as saltires.}
\label{f:tidc}
\end{center}
\end{figure*}

With regard to a generic configuration, as the one depicted in
Fig.\,\ref{f:tidi}, it can be seen $\gamma<\eta$ and $\gamma>\eta$ for bound
and (partially) unbound embedded sphere, respectively.
The point, $(\gamma,\eta)=(1,1)$, relates to the
configuration where the centre of the embedding sphere lies on the surface of
the embedded sphere, as depicted in Fig.\,\ref{f:tidc}.
The point, $(\gamma,\eta)=(0,0)$, relates to the
configuration where the embedding sphere and the (no longer) embedded sphere
are infinitely distant.

The subdomain,
$0\le\gamma<1$, makes a necessary condition for the occurrence of the tidal
radius, in that the gravitational force exerted on the point,
${\sf P}$, from either subsystem acts along the same direction but on
opposite sides, as depicted in Fig.\,\ref{f:tidi}.    The special case,
$\gamma=1$, relates to a null gravitational force on ${\sf P}$ from
the embedding sphere.   The subdomain, $\gamma>1$, implies
$R_{\rm C}-a_{\rm C}<0$ and, in consequence, a mass enclosed within a negative
volume.   Accordingly, further considerations shall be restricted to the
subdomain, $0\le\gamma\le1$, where the physical meaning is preserved, which
implies $0\le\eta\le1$.

\subsection{Application to globular clusters within galaxies}
\label{gcwg}
\subsubsection{One-component embedding density profiles}
\label{1cdp}

As a guiding example, let the embedding sphere be considered with fixed
truncation radius, $a_{\rm G}=100$ kpc,
mass, $M_{\rm G}=M_{\rm G}(a_{\rm G})=10^{12}m_\odot$,
concentration, $\Xi=10$,
and density profiles selected among those
listed in Table \ref{t:gadp} which are currently used for fitting to galaxies
and/or dark matter haloes.   Similarly, let the embedded sphere be considered
with fixed truncation radius, $a_{\rm C}=30$ pc, and mass,
$M_{\rm C}=M_{\rm C}(a_{\rm C})=10^5m_\odot$, regardless of the density
profile.   The values of fractional mass,
$\nu_{\rm mas}=M_{\rm G}/M_{\rm G}^\dagger$,
scaling mass, $M_{\rm G}^\dagger$, scaling density, $\rho_{\rm G}^\dagger$,
fractional distance related to tidal radius,
$(\xi^\ast)^\mp=(\xi_{\rm C}^\ast)^\mp-(\delta_{\rm C}^\ast)^\mp=
[(R_{\rm C}^\ast)^\mp-(a_{\rm C}^\ast)^\mp]/r_{\rm G}^\dagger$, are
listed in Table \ref{t:prode}.   For H density profiles $(\xi^\ast)^-=0$,
which implies $(\gamma^\ast)^-=1$, hence $(R_{\rm C}^\ast)^-\to+\infty$,
$(a_{\rm C}^\ast)^-\to+\infty$.
\begin{table}
\caption{Fractional mass, $\nu_{\rm mas}=M_{\rm G}/M_{\rm G}^\dagger$,
scaling mass, $M_{\rm G}^\dagger$ (unit $10^{10}m_\odot$), scaling density,
$\rho_{\rm G}^\dagger$ (unit $10^{10}m_\odot/{\rm kpc}^3$),
fractional distance related to tidal radius,
$(\xi^\ast)^\mp=(\xi_{\rm C}^\ast)^\mp-(\delta_{\rm C}^\ast)^\mp=
[(R_{\rm C}^\ast)^\mp-(a_{\rm C}^\ast)^\mp]/r_{\rm G}^\dagger$,
for embedding one-component density profiles ($p$) listed in Table
\ref{t:gadp} and
currently used for the description of galaxies and/or dark matter haloes.
In any case, the following parameters remain unchanged:
truncation radius, $a_{\rm G}=100$ kpc; total
mass, $M_{\rm G}=10^{12}m_\odot$; concentration, $\Xi=10$.
With regard to the embedded sphere, the truncation radius and the total mass
are kept fixed to $a_{\rm C}=30$ pc and $M_{\rm C}=10^5m_\odot$, respectively.
For H density profiles $(\xi^\ast)^-$ is null, which implies both
$(R_{\rm C}^\ast)^-$ and $(a_{\rm C}^\ast)^-$ extend to infinite.
See text for further details.}
\label{t:prode}
\begin{center}
\begin{tabular}{clllll} \hline
\multicolumn{1}{c}{$p$}
&\multicolumn{1}{c}{$\nu_{\rm mas}$}
&\multicolumn{1}{c}{$M^\dagger_{\rm G}$}
&\multicolumn{1}{c}{$\rho^\dagger_{\rm G}$}
&\multicolumn{1}{c}{$(\xi^\ast)^-$}
&\multicolumn{1}{c}{$(\xi^\ast)^+$} \\
\noalign{\smallskip}
P & 5.5730D+0 & 1.7943D+1 & 4.2837D$-$3 & 1.0949D$-$2 & 9.4789D+0 \\
H & 9.9174D+0 & 1.0083D+1 & 2.4072D$-$3 & 0           & 9.4355D+0 \\
I & 5.1173D+0 & 1.9541D+0 & 4.6652D$-$4 & 2.9944D$-$1 & 8.8055D+0 \\
S & 1.6998D+1 & 5.8832D+0 & 1.4045D$-$3 & 6.7043D$-$2 & 9.3217D+0 \\
B & 1.9406D+1 & 5.1531D+0 & 1.2302D$-$3 & 5.6281D$-$2 & 9.2917D+0 \\
\noalign{\smallskip}      
\hline                                                       
\end{tabular}                                                
\end{center}                                                 
\end{table}                                                  

\begin{figure*}[t]
\begin{center}
\includegraphics[scale=0.8]{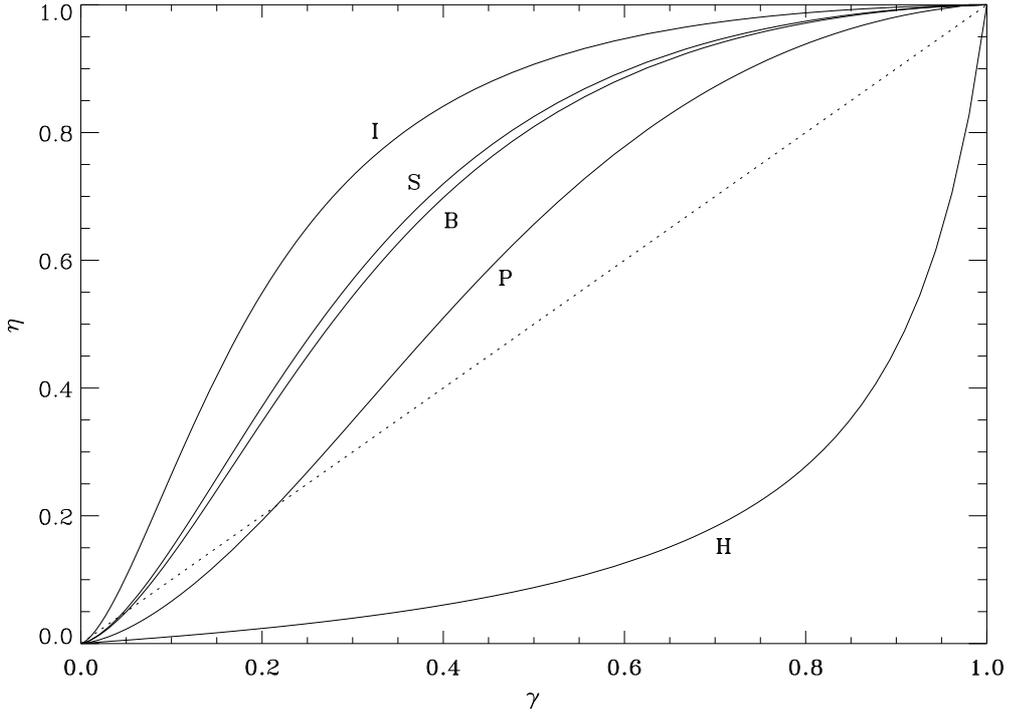}
\caption[OKND]{The instantaneous tidal radius, $\eta=\mu/(\mu+1)$, vs. the
inverse fractional distance, $\gamma=a_{\rm C}/R_{\rm C}$, for an embedded sphere
of mass, $M_{\rm C}=10^5m_\odot$, truncation radius, $a_{\rm C}=30$ pc, and an
embedding sphere of mass, $M_{\rm G}=10^{12}m_\odot$,
truncation radius,
$a_{\rm G}=100$ kpc, concentration, $\Xi=10$,
density profiles listed in Table
\ref{t:prode} and captioned as: B - Burkert (1995); H - Hernquist (1990); I - 
Begemann et al. (1991); P - Plummer (1911); S - Spano et al. (2008).   The
locus, $\eta=\gamma$, is also shown as a dotted line, above which
$(\gamma<\eta)$ the embedded sphere is considered bound according to the
assumed formulation of tidal radius.  See text for further details.}
\label{f:mare}
\end{center}
\end{figure*}

\begin{figure*}[t]
\begin{center}
\includegraphics[scale=0.8]{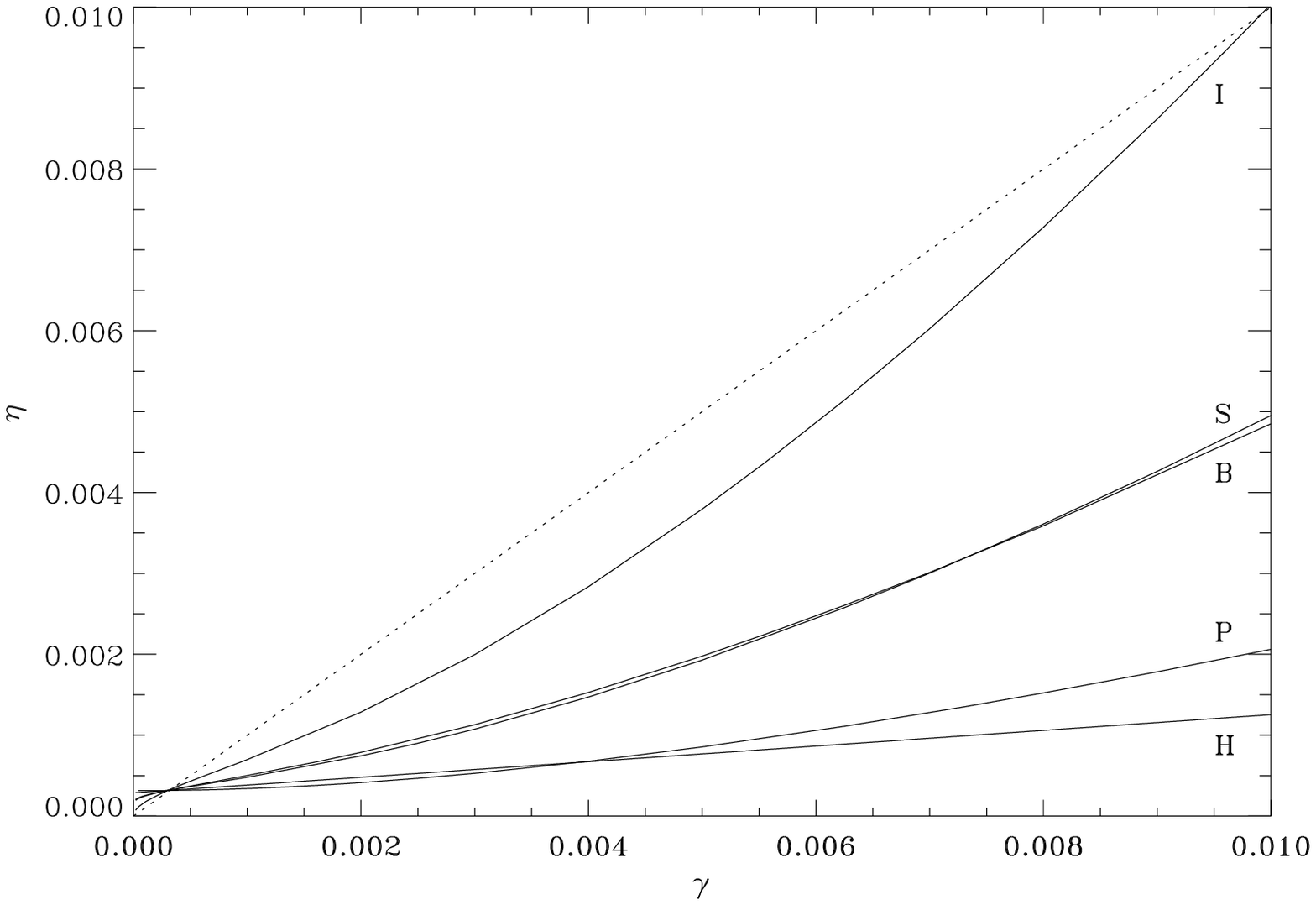}
\caption[OKND]{Zoom of Fig.\,\ref{f:mare} near the origin.   The
truncation radius of the embedding sphere is attained just after the
intersection of related curve with the dotted straight line, $\eta=\gamma$, 
i.e. just after the tidal radius and the truncation radius of the embedded
sphere coincide.   The continuation of each curve outside the truncation 
radius of the embedding sphere, assuming the continuation of the density
profile, is also shown.}
\label{f:mari}
\end{center}
\end{figure*}

The trend of $\eta$
vs. $\gamma$ for the cases listed in Table \ref{t:prode} is plotted in
Fig.\,\ref{f:mare}, while the neighbourhood of the origin is zoomed in
Fig.\,\ref{f:mari}, where density profiles are continued outside the
truncation radius.
%

An inspection of Figs.\,\ref{f:mare} and \ref{f:mari} shows that, in the light
of the assumed formulation of tidal radius, the embedded sphere is bound
provided close enough to $(\gamma\appgeq0.01$-$0.2)$ and far enough from
$(\gamma\appleq0.0003)$ the centre of the embedding sphere.   Conversely,
the embedded sphere is (partially) unbound provided sufficiently far from both
the centre and the surface $(0.0003\appleq\gamma\appleq0.01$-$0.2)$ of the
embedding sphere.   The sole exception is from H density profiles, which
implies (partially) unbound embedded spheres up to $\gamma\approx0.0003$ i.e.
close to the truncation radius.

\subsubsection{Two-component embedding density profiles}
\label{22cdp}

As a guiding example, let the embedding two-component sphere be considered
with fixed
truncation radius, $a_{i,{\rm G}}=30$ kpc, $a_{j,{\rm G}}=100$ kpc;
mass,  $M_{i,{\rm G}}=10^{11}m_\odot$,  $M_{j,{\rm G}}=10^{12}m_\odot$;
concentration, $\Xi_i=\Xi_j=10$;
and density profiles as specified in Section \ref{2cdp}.
Similarly, let the embedded sphere be considered
with fixed truncation radius, $a_{\rm C}=30$ pc, and mass,
$M_{\rm C}=10^5m_\odot$, regardless of the density
profile.   The values of fractional mass,
$\nu_{u,{\rm mas}}=M_{u,{\rm G}}/M_{u,{\rm G}}^\dagger$, 
scaling mass, $M_{u,{\rm G}}^\dagger$, scaling density,
$\rho_{u,{\rm G}}^\dagger$, fractional distance related to tidal radius,
$(\xi_u^\ast)^\mp=(\xi_{u,{\rm C}}^\ast)^\mp-(\delta_{u,{\rm C}}^\ast)^\mp=
[(R_{\rm C}^\ast)^\mp-(a_{\rm C}^\ast)^\mp]/r_{u,{\rm G}}^\dagger$, $u=i,j$,
are listed in Table \ref{t:prod2}.   For H density profiles $(\xi^\ast)^-=0$,
which implies $(\gamma^\ast)^-=1$, hence $(R_{\rm C}^\ast)^-\to+\infty$,
$(a_{\rm C}^\ast)^-\to+\infty$, and, in
addition, $(\xi_j)^+>\Xi_j$ i.e. $a_{\rm C}=a_{\rm C}^\ast$ provided
$R_{\rm C}\appgeq a_{j,{\rm G}}$.
\begin{table}
\caption{Fractional mass,
$\nu_{u,{\rm mas}}=M_{u,{\rm G}}/M_{u,{\rm G}}^\dagger$,
scaling mass, $M_{u,{\rm G}}^\dagger$ (unit $10^{10}m_\odot$), scaling
density, $\rho_{u,{\rm G}}^\dagger$ (unit $10^{10}m_\odot/{\rm kpc}^3$),
fractional distance related to tidal radius,
$(\xi_u^\ast)^\mp=(\xi_{u,{\rm C}}^\ast)^\mp-(\delta_{u,{\rm C}}^\ast)^\mp=
[(R_{\rm C}^\ast)^\mp-(a_{\rm C}^\ast)^\mp]/r_{u,{\rm G}}^\dagger$,
for embedding two-component density profiles ($ij$) mentioned in
Section \ref{2cdp} and usable
for the description of galaxies within dark matter haloes.
In any case, the following parameters remain unchanged:
truncation radius, $a_{i,{\rm G}}=30$ kpc, $a_{j,{\rm G}}=100$ kpc; total
mass, $M_{i,{\rm G}}=10^{11}m_\odot$, $M_{j,{\rm G}}=10^{12}m_\odot$;
concentration, $\Xi_i=\Xi_j=10$.
With regard to the embedded sphere, the truncation radius and the total mass
are kept fixed to $a_{\rm C}=30$ pc and $M_{\rm C}=10^5m_\odot$, respectively.
For each case, upper and lower lines relate to the inner ($u=i$, baryonic) and
outer ($u=j$, nonbaryonic) embedding subsystem.
See text for further details.}
\label{t:prod2}
\begin{center}
\begin{tabular}{clllll} \hline
\multicolumn{1}{c}{$u$}
&\multicolumn{1}{c}{$\nu_{u,{\rm mas}}$}
&\multicolumn{1}{c}{$M^\dagger_{u,{\rm G}}$}
&\multicolumn{1}{c}{$\rho^\dagger_{u,{\rm G}}$}
&\multicolumn{1}{c}{$(\xi_u^\ast)^-$}
&\multicolumn{1}{c}{$(\xi_u^\ast)^+$} \\
\noalign{\smallskip}
\hline\noalign{\smallskip}
P & 5.5730D+0 & 1.7943D+0 & 1.5866D$-$2 & 9.7519D$-$3 & 3.3024D+1 \\
I & 5.1173D+1 & 1.9541D+0 & 4.6652D$-$4 & 2.9256D$-$3 & 9.9073D+0 \\
\hline\noalign{\smallskip}
H & 9.9174D+0 & 1.0083D+0 & 8.9156D$-$3 & 0           & 3.3426D+1 \\
I & 5.1173D+1 & 1.9541D+0 & 4.6652D$-$4 & 0           & 1.0028D+1 \\
\hline\noalign{\smallskip}
P & 5.5730D+0 & 1.7943D+0 & 1.5866D$-$2 & 9.4355D$-$3 & 3.3145D+1 \\
S & 1.6998D+1 & 5.8832D+0 & 1.4045D$-$3 & 2.8306D$-$3 & 9.9434D+0 \\
\hline\noalign{\smallskip}
H & 9.9174D+0 & 1.0083D+0 & 8.9156D$-$3 & 0           & 3.3390D+1 \\
S & 1.6998D+1 & 5.8832D+0 & 1.4045D$-$3 & 0           & 1.0017D+1 \\
\noalign{\smallskip}      
\hline                                                       
\end{tabular}                                                
\end{center}                                                 
\end{table}                                                  

\begin{figure*}[t]
\begin{center}
\includegraphics[scale=0.8]{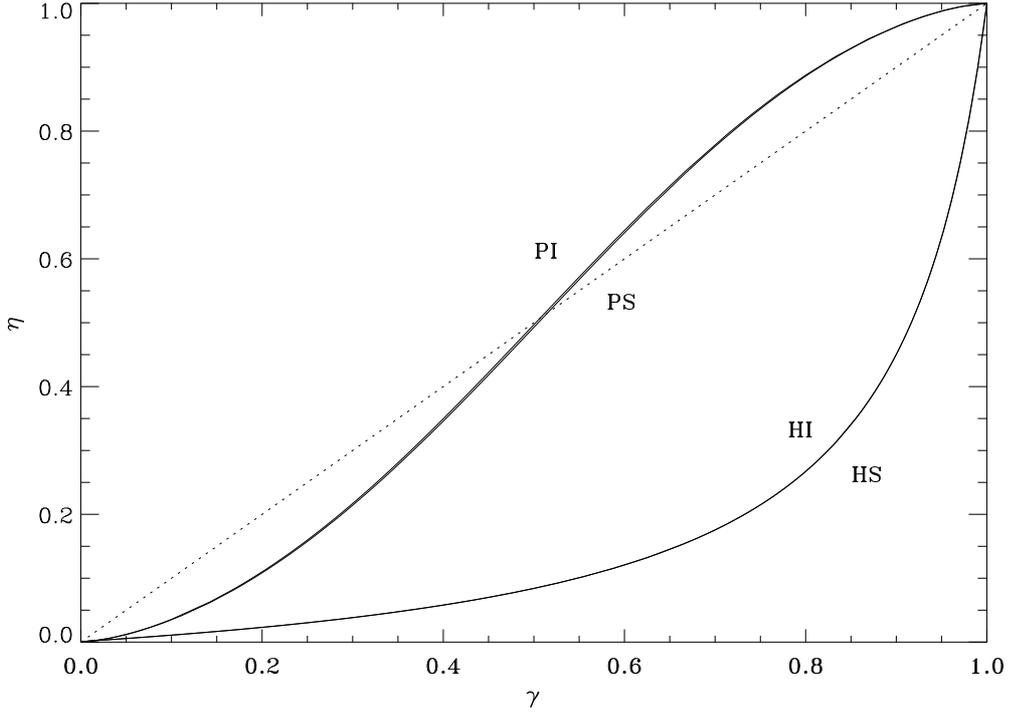}
\caption[OKND]{The instantaneous tidal radius, $\eta=\mu/(\mu+1)$, vs. the
inverse fractional distance, $\gamma=a_{\rm C}/R_{\rm C}$, for an embedded
sphere of mass, $M_{\rm C}=10^5m_\odot$, truncation radius, $a_{\rm C}=30$ pc,
and an embedding two-component sphere of mass, $M_{i,{\rm G}}=10^{11}m_\odot$,
$M_{j,{\rm G}}=10^{12}m_\odot$, truncation radius, $a_{i,{\rm G}}=30$ kpc,
$a_{j,{\rm G}}=100$ kpc, concentration, $\Xi_i=\Xi_j=10$, density profiles,
$ij=$ PI, PS, HI, HS, captioned as in Fig.\,\ref{f:mare}.   Curves related to
PI, PS; HI, HS; density profiles, respectively, are nearly coincident and
cannot be resolved on the plane of the figure.   The
locus, $\eta=\gamma$, is also shown as a dotted line, above which
$(\gamma<\eta)$ the embedded sphere is considered bound according to the
assumed formulation of tidal radius.  See text for further details.}
\label{f:mae2}
\end{center}
\end{figure*}

\begin{figure*}[t]
\begin{center}
\includegraphics[scale=0.8]{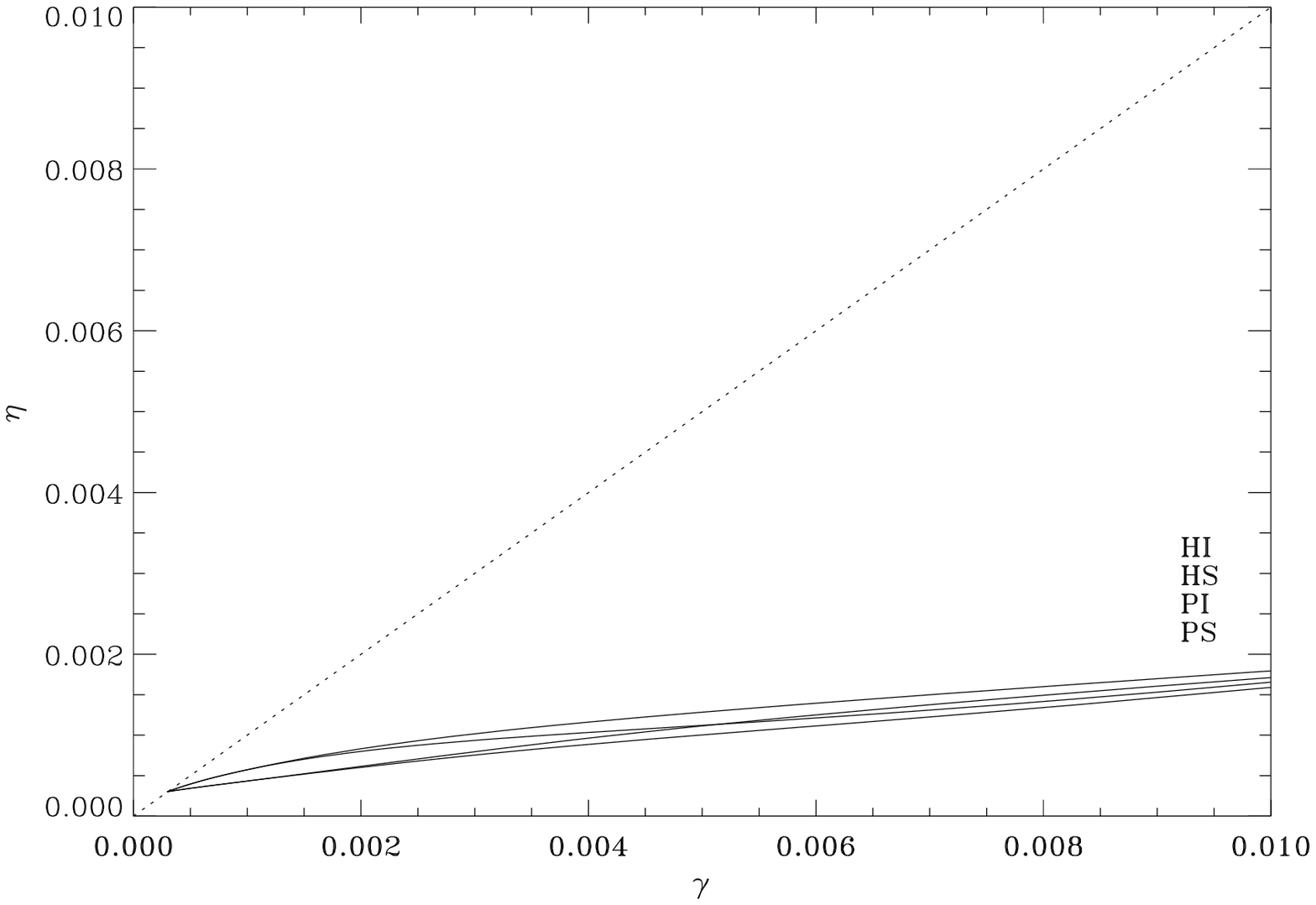}
\caption[OKND]{Zoom of Fig.\,\ref{f:mae2} near the origin.   The ending point 
of each curve (left) relates to the truncation radius of the outer embedding
sphere $(j=$ I,S), where $\gamma=0.0003$.   The scale has been left unchanged
with respect to Fig.\,\ref{f:mari} for better comparison.}
\label{f:mai2}
\end{center}
\end{figure*}

The trend of $\eta$
vs. $\gamma$ for the cases listed in Table \ref{t:prode} is plotted in
Fig.\,\ref{f:mae2}, while the neighbourhood of the origin is zoomed in
Fig.\,\ref{f:mai2}, where density profiles end at the truncation radius of the
outer embedding sphere, related to $\gamma=0.0003$.
%

An inspection of Figs.\,\ref{f:mae2} and \ref{f:mai2} shows that, in the light
of the assumed formulation of tidal radius, the embedded sphere is bound for
PI and PS density profiles, provided it is close enough to
$(\gamma\appgeq0.5)$ and far enough from
$(\gamma\appleq0.0003)$ the centre of the embedding sphere.   Conversely,
the embedded sphere is (partially) unbound provided sufficiently far from both
the centre and the surface $(0.0003\appleq\gamma\appleq0.5)$ of the outer
embedding sphere.   On the other hand, HI and HS density profiles
imply (partially) unbound embedded spheres up to $\gamma\appgeq0.0003$ i.e.
for galactocentric distances slightly exceeding the truncation radius of the
outer embedding sphere, $R_{\rm C}\appgeq a_{j,{\rm G}}$.
The curves are mainly depending on the inner density profile ($i=$ P,H), while
the effect of the outer density profile ($j=$ I,S) can be neglected to a
good extent.

The occurrence of an extremum point (maximum absolute value) in the gravitational
acceleration profile is also determined by the inner subsystem for the cases
under discussion.   With regard to PI and PS density profiles, the
dimensionless effective radius, via Eqs.\,(\ref{eq:gePI}) and
(\ref{eq:gePS}), respectively, takes place at
$(\xi_{\rm eff})_{\rm P}=0.716\,584\,739$ or equivalently
$(\xi_{\rm eff})_{\rm I}=0.214\,975\,422$ and
$(\xi_{\rm eff})_{\rm P}=0.663\,475\,075$ or equivalently
$(\xi_{\rm eff})_{\rm S}=0.199\,042\,522$, respectively.    With regard to HI
and HS density profiles, Eqs.\,(\ref{eq:geHI}) and (\ref{eq:geHS}) show no
solution, which implies no extremum point in the gravitational acceleration.

\section{Discussion}
\label{disc}

With regard to tidal effects, the above results are grounded on the definition
of tidal radius, which has been selected for
reasons of simplicity.   Though using different definitions
implies different results (e.g., Brosche et al. 1999; Caimmi and Secco 2003),
a similar trend is expected.   According to the formulation of the current
paper, a test particle on the surface of the embedded sphere remains no longer
bound if the gravitational force from the embedded sphere is counterbalanced
by the gravitational force from the embedding sphere.   To this respect, small
perturbations suffice to make a test particle be lost from the embedded
sphere.
To gain more insight, let the embedding sphere, the embedded sphere, the test
particle, be conceived as a galaxy (in particular, the Galaxy), a globular
cluster (GC), a (long-lived) star, respectively.

By definition, all stars lying on GC boundaries necessarily exhibit null
radial velocity or, in other words, the stars under consideration are at the
apocentre of their orbits.   Among elliptic orbits with different
eccentricities and equal major axis, the pendulum orbit has the lowest energy,
due to a null tangential velocity on the apocentre%
\footnote{
It is worth noticing the major axis of the pendulum orbit doubles the major
axis of the Keplerian orbit with unit eccentricity, for fixed apocentre.
}.
Stars with nonzero tangential velocity on GC boundaries should be less bound
as in presence of centrifugal force.   On the other hand the centrifugal
force, due to GC orbital motion within the embedding galaxy, has no influence
on the above mentioned gravitational balance.

Then the definition of tidal
radius, assumed in the current paper, relates to a necessary condition.  More
specifically, bound GCs imply $\gamma<\eta$ but the reverse could not hold.
With this caveat in mind, it shall be supposed in the following the condition
is also sufficient i.e. $\gamma<\eta$ implies bound GCs.

For a galaxy of assigned mass, truncation radius, concentration, and for a
specified GC, the trend of $\eta$ vs. $\gamma$ depends on the galactic density
profile.   An inspection of Figs.\,\ref{f:mare}-\ref{f:mari} shows a far more
extended stability region, $\gamma<\eta$, for density profiles (P, I, S, B)
where the gravitational acceleration has an extremum point, with respect to
density profiles (H) where the gravitational acceleration is monotonically
increasing in absolute value.   Different GCs imply different masses,
$M_{\rm C}$, and
truncation radii, $a_{\rm C}$, which translates into different curves on the
$({\sf O}\gamma\eta)$ plane.

Restricting to the Galaxy with GC subsystem included, let the following values
be assumed: mass, $M_{\rm G}=5\cdot10^{10}m_\odot$; truncation radius,
$a_{\rm G}=125$ kpc, concentration, $\Xi=10$.   Let a GC sample $(N=16)$
studied in an earlier attempt (Brosche et al. 1999) be considered, with the
addition of a further element (Pal5) for which different masses can be
inferred (e.g., Caimmi and Secco, 2003).   The GC subsystem (thick disk, old
halo, young halo), S, observed radius, $r_{\rm C}$, galactocentric distance,
$R_{\rm C}$, mass logarithm, $\log(M_{\rm C}/m_\odot)$, taken from the parent
paper (Brosche et al. 1999), are listed in Table \ref{t:glob} together with
the inferred inverse fractional distance, $\gamma=r_{\rm C}/R_{\rm C}$.   The
inferred fractional instantaneous tidal radius, $\eta=\mu/(1+\mu)$, for
different density profiles among those listed in Table \ref{t:gadp}, is shown
in Table \ref{t:glo1}, where the inverse fractional distance is repeated for
better comparison with related plots.
\begin{table}
\caption{Parameters of globular clusters studied in an earlier paper (Brosche
et al. 1999).
An additional cluster, Pal5, considered in a subsequent paper
(Caimmi and Secco 2003) is also included for different mass values.   Column
caption: 1 - name (NGC or Pal); 2 - subsystem (A - [Fe/H] $>-$1, thick disk;
B - old halo; C - young halo); 3 - observed radius, $r_{\rm C}/$pc; 4 -
galactocentric distance, $R_{\rm C}/$kpc; 5 - decimal logarithm of mass,
$\log\phi=\log(M_{\rm C}/m_\odot)$; 6 - inverse fractional distance,
$\gamma=r_{\rm C}/R_{\rm C}$.}
\label{t:glob}
\begin{center}
\begin{tabular}{lcrrrr} \hline
\multicolumn{1}{c}{NGC}
&\multicolumn{1}{c}{S}
&\multicolumn{1}{c}{$\frac{r_{\rm C}}{\rm pc}$}
&\multicolumn{1}{c}{$\frac{R_{\rm C}}{\rm kpc}$}
&\multicolumn{1}{c}{$\log\phi$}
&\multicolumn{1}{c}{$10^3\gamma$} \\
\noalign{\smallskip}
     &   &                 &      &      &           \\
     &   &                 &      &      &           \\
0104 & A &  50.7           &  7.4 & 6.16 &  6.85135  \\ 
0362 & C &  35.7           &  9.3 & 5.75 &  3.83871  \\ 
4147 & C &  34.5           & 21.3 & 4.85 &  1.61972  \\
5024 & C & 119.3           & 18.8 & 5.91 &  6.34574  \\
5272 & C & 103.0           & 12.2 & 5.95 &  8.44262  \\ 
5466 & C & 101.4           & 17.2 & 5.23 &  5.89535  \\
5904 & C &  63.0           &  6.2 & 5.91 & 10.16129  \\ 
6205 & B &  55.4           &  8.7 & 5.81 &  6.36782  \\ 
6218 & B &  21.6           &  4.5 & 5.32 &  4.80000  \\ 
6254 & B &  27.0           &  4.6 & 5.38 &  5.86957  \\ 
6341 & B &  35.0           &  9.6 & 5.67 &  3.64583  \\ 
6779 & B &  25.0           &  9.7 & 5.34 &  2.57732  \\
6838 & A &  10.1           &  6.7 & 4.61 &  1.50746  \\
6934 & C &  37.5           & 14.3 & 5.39 &  2.62238  \\
7078 & C &  65.7           & 10.4 & 6.05 &  6.31731  \\ 
7089 & B &  71.1           & 10.4 & 6.00 &  6.83654  \\ 
     &   &                 &      &      &           \\
Pal5 & C &  20\phantom{.2} & 18.6 & 3.78 &  1.07527  \\
     &   &  20\phantom{.2} & 18.6 & 3.65 &  1.07527  \\
     &   &  20\phantom{.2} & 18.6 & 3.15 &  1.07527  \\
     &   &  20\phantom{.2} & 18.6 & 2.98 &  1.07527  \\
\noalign{\smallskip}      
\hline                                                       
\end{tabular}                                                
\end{center}                                                 
\end{table}                                                  
\begin{table}
\caption{Fractional instantaneous tidal radius, $\eta=\mu/(1+\mu)$, inferred
for globular clusters listed in Table \ref{t:glob}, embedded within a
one-component sphere of mass, $M_{\rm G}=5\cdot10^{10}m_\odot$; truncation
radius; $a_{\rm G}=125$ kpc; concentration, $\Xi=10$; one-component density
profiles among those listed in Table \ref{t:gadp}.   The inverse fractional
distance, $\gamma=r_{\rm C}/R_{\rm C}$, is also repeated for better comparison
with related plots.}
\label{t:glo1}
\begin{center}
\begin{tabular}{lrrrrrr} \hline
\multicolumn{1}{c}{NGC}
&\multicolumn{1}{c}{$10^3\gamma$}
&\multicolumn{5}{c}{$10^3\eta$} \\
\noalign{\smallskip}
     &          &  P~~~~~~ & I~~~~~~~ & H~~~~~~~ & S~~~~~~~ & B~~~~~~~  \\
     &          &          &          &          &          &           \\
0104 &  6.85135 & 14.57470 & 61.99897 & 13.02944 & 32.40667 & 33.18122  \\ 
0362 &  3.83871 &  7.19259 & 29.56671 &  7.11140 & 15.52356 & 16.11275  \\ 
4147 &  1.61972 &  1.47333 &  4.25334 &  1.71465 &  2.53435 &  2.68024  \\
5024 &  6.34574 &  5.25419 & 16.16833 &  6.08087 &  9.41380 &  9.95776  \\
5272 &  8.44262 &  7.17412 & 26.91316 &  7.74374 & 14.58642 & 15.31472  \\ 
5466 &  5.89535 &  2.51561 &  8.15645 &  2.89185 &  4.64889 &  4.91754  \\
5904 & 10.16129 & 13.49411 & 59.55797 & 11.00866 & 30.72825 & 31.10789  \\ 
6205 &  6.36782 &  8.24594 & 34.44840 &  7.92721 & 18.00489 & 18.61784  \\ 
6218 &  4.80000 & 10.25214 & 47.63253 &  6.99596 & 24.08752 & 23.88570  \\ 
6254 &  5.86957 & 10.68040 & 49.44294 &  7.37929 & 25.04429 & 24.86855  \\ 
6341 &  3.64583 &  6.36749 & 25.99165 &  6.37330 & 13.66944 & 14.21220  \\ 
6779 &  2.57732 &  4.31703 & 17.67155 &  4.33957 &  9.26801 &  9.64317  \\
6838 &  1.50746 &  2.75645 & 12.44064 &  2.34830 &  6.30028 &  6.41645  \\
6934 &  2.62238 &  3.36082 & 11.87662 &  3.76543 &  6.55913 &  6.92043  \\
7078 &  6.31731 &  9.15972 & 36.16933 &  9.42575 & 19.28644 & 20.12065  \\ 
7089 &  6.83654 &  8.65576 & 34.23661 &  8.90570 & 18.23766 & 19.02693  \\ 
     &          &          &          &          &          &           \\
Pal5 &  1.07527 &  0.04547 &  1.41764 &  0.52651 &  0.81934 &  0.86712  \\
     &  1.07527 &  0.03938 &  1.22795 &  0.45600 &  0.70965 &  0.75104  \\
     &  1.07527 &  0.02205 &  0.68773 &  0.25530 &  0.39736 &  0.42054  \\
     &  1.07527 &  0.01811 &  0.56487 &  0.20968 &  0.32636 &  0.34540  \\
\noalign{\smallskip}      
\hline                                                       
\end{tabular}                                                
\end{center}                                                 
\end{table}                                                  
Concerning Pal5, calculations were performed with regard to four possible mass
values: for further details and references, an interested reader is
addressed to the parent paper (Caimmi and Secco 2003).

\begin{figure*}[t]
\begin{center}
\includegraphics[scale=0.8]{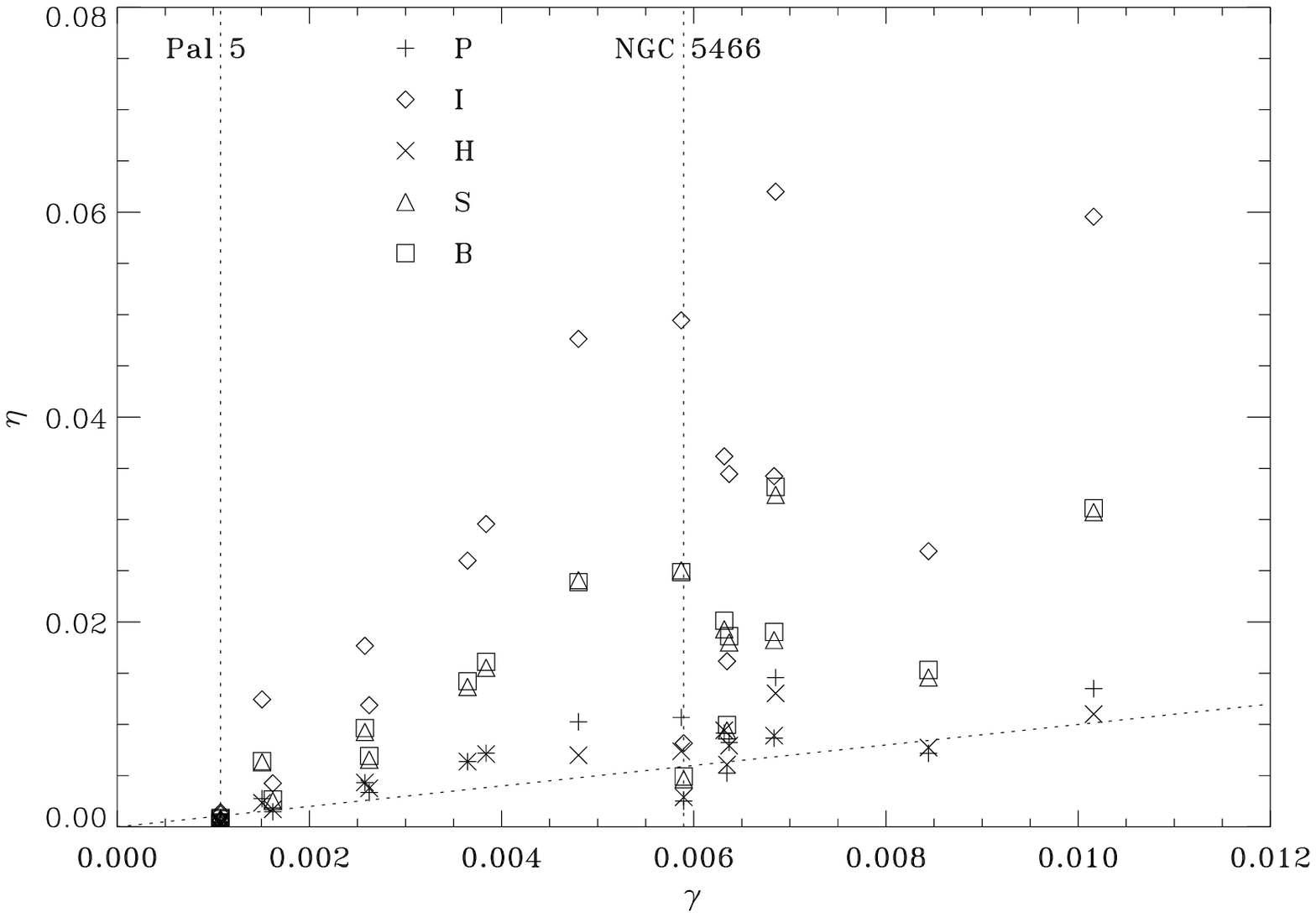}
\caption[OKND]{Location of globular clusters, listed in Table \ref{t:glo1}, on
the $({\sf O}\gamma\eta)$ plane, for an embedding sphere of mass,
$M_{\rm G}=5\cdot10^{10}m_\odot$, truncation radius, $a_{\rm G}=125$ kpc, 
concentration, $\Xi=10$, density profiles among those listed in Table
\ref{t:gadp}, captioned as in Fig.\,\ref{f:mare}.   The locus, $\eta=\gamma$,
is also shown as an inclined dotted
line, above which $(\gamma<\eta)$ globular clusters are bound according to the
assumed formulation of tidal tadius.   The position of NGC 5466 and Pal5,
which exhibit noticeable tidal effects, are marked by vertical dotted lines.  See
text for further details.}
\label{f:glob}
\end{center}
\end{figure*}

The location of GCs listed in Table \ref{t:glo1} on the $({\sf O}\gamma\eta)$
plane is shown in Fig.\,\ref{f:glob} for different Galactic density profiles
among those
listed in Table \ref{t:gadp}.   The region where GCs are bound, according to
the assumed formulation of tidal radius, lies above the dotted straight line,
$\eta=\gamma$.
The inverse fractional distance, $\gamma$, related to NGC 5466 and Pal5, is
marked by a dotted vertical line.

The above mentioned GCs appear (partially)
unbound for all considered density profiles, with the possible exception of
I.   In fact, Pal5 is known to be experiencing progressive disruption via
tidal shoks during disk passages (e.g., Odenkirchen et al. 2002; Dehnen et al.
2004) and NGC 5466 shows a tidal stream (Grillmair and Johnson 2006), as
inferred in an earlier attempt for different formulations of tidal radius
(Caimmi and Secco 2003).   Remaining GCs are placed inside the stability
region, $\gamma<\eta$, with a restricted number close to the boundary,
$\gamma=\eta$, for all considered density profiles.

The key parameter, on which the above results mainly depend, is the mass of
the Galaxy.   In fact, a longer Galactic truncation radius implies less amount
of mass acting on a selected GC and vice versa.   Accordingly, the tidal
force is proportional to the Galactic mass and (for unchanged Galactic mass)
inversely proportional to the Galactic truncation radius, as expected.

With regard to the results listed in Table \ref{t:glo1} and plotted in
Fig.\,\ref{f:glob}, the Galactic mass is restricted to stars and gas leaving
aside nonbaryonic dark matter.   To this respect, the presence of additional
mass makes the inverse fractional distance, $\gamma$, unchanged by definition
via  Eq.\,(\ref{eq:gamu}), while the fractional tidal radius, $\eta=\eta^+$,
by definition via  Eq.\,(\ref{eq:istr}), maintains the same formal expression
where the mass ratio, $\mu(\xi_{\rm C}-\delta_{\rm C})$, reads:
\begin{lefteqnarray}
\label{eq:muto}
&& \mu(\xi_{\rm C}-\delta_{\rm C})=\left[\frac{1+\zeta_{\rm C}}
{1+\zeta_{\rm G}(\xi_{\rm C}-\delta_{\rm C})}\right]^{1/2}
\left[\frac{M_{\rm C}}{M_{\rm G}(\xi_{\rm C}-\delta_{\rm C})}\right]^{1/2}~~;
\end{lefteqnarray}
where masses are still restricted to baryonic matter and $\zeta$ is the ratio
between nonbaryonic and baryonic mass for a selected subsystem.   A constant
ratio, $\zeta_{\rm G}(\xi_{\rm C}-\delta_{\rm C})=\zeta_{\rm C}=\zeta$,
implies the results are left unchanged in that Eq.\,(\ref{eq:muto}) reduces to
(\ref{eq:mutr}).

\begin{figure*}[t]
\begin{center}
\includegraphics[scale=0.8]{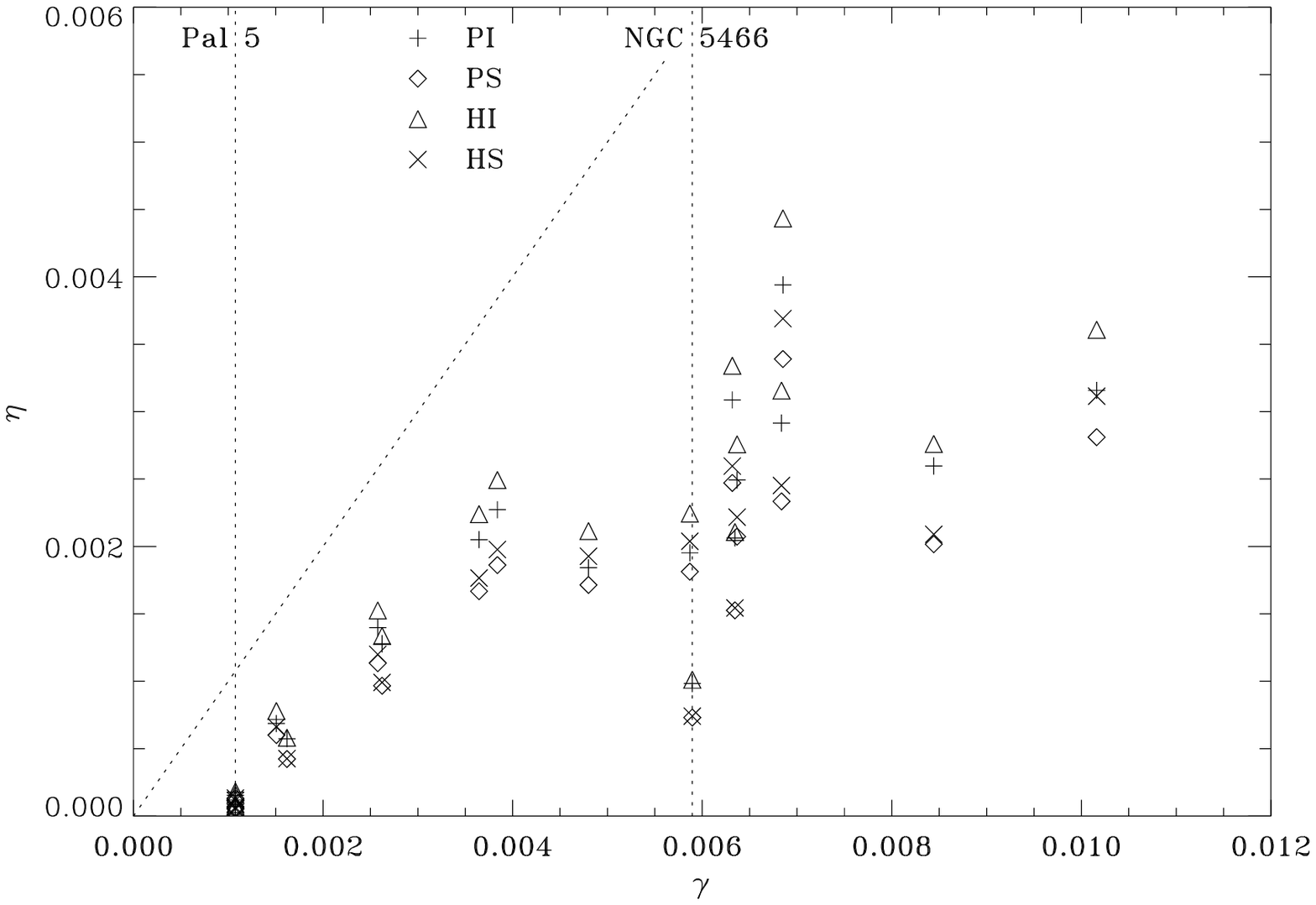}
\caption[OKND]{Location of globular clusters, listed in Table \ref{t:glo2}, on
the $({\sf O}\gamma\eta)$ plane, for a two-component embedding sphere of mass,
$M_{i,{\rm G}}=10^{11}m_\odot$, $M_{j,{\rm G}}=10^{12}m_\odot$; truncation 
radius, $a_{i,{\rm G}}=30$ kpc, $a_{j,{\rm G}}=100$ kpc; concentration, 
$\Xi_i=\Xi_j=10$; two-component density profiles considered in  Subsection  
\ref{22cdp}, captioned as in Fig.\,\ref{f:mae2}.   The locus, $\eta=\gamma$,
is also shown as an inclined dotted
line, above which $(\gamma<\eta)$ globular clusters are bound according to the
assumed formulation of tidal tadius.   The position of NGC 5466 and Pal5,
which exhibit noticeable tidal effects, are marked by vertical dotted lines.  See
text for further details.}
\label{f:glo2}
\end{center}
\end{figure*}

The above procedure is repeated for two-component embedding spheres considered
in Subsection \ref{22cdp} and the results are listed in Table \ref{t:glo2} and
plotted in Fig.\,\ref{f:glo2}. 
\begin{table}
\caption{Fractional instantaneous tidal radius, $\eta=\mu/(1+\mu)$, inferred
for globular clusters listed in Table \ref{t:glob}, embedded within a
two-component $(ij)$ sphere of mass, 
$M_{i,{\rm G}}=10^{11}m_\odot$, $M_{j,{\rm G}}=10^{12}m_\odot$; truncation 
radius, $a_{i,{\rm G}}=30$ kpc, $a_{j,{\rm G}}=100$ kpc; concentration, 
$\Xi_i=\Xi_j=10$; two-component density profiles considered in  Subsection  
\ref{22cdp}.   The inverse fractional distance, $\gamma=r_{\rm C}/R_{\rm C}$,
is also repeated for better comparison with related plots.}
\label{t:glo2}
\begin{center}
\begin{tabular}{lrrrrr} \hline
\multicolumn{1}{c}{NGC}
&\multicolumn{1}{c}{$10^3\gamma$}
&\multicolumn{4}{c}{$10^3\eta$} \\
\noalign{\smallskip}
     &          & PI~~~~~ & PS~~~~~ & HI~~~~~ & HS~~~~~  \\
     &          &         &         &         &          \\
0104 &  6.85135 & 3.93932 & 3.39026 & 4.43380 & 3.69007  \\ 
0362 &  3.83871 & 2.27233 & 1.86155 & 2.49265 & 1.97695  \\ 
4147 &  1.61972 & 0.57317 & 0.42316 & 0.58079 & 0.42620  \\
5024 &  6.34574 & 2.06309 & 1.52597 & 2.10705 & 1.54349  \\
5272 &  8.44262 & 2.59695 & 2.01639 & 2.76012 & 2.08992  \\ 
5466 &  5.89535 & 0.98317 & 0.73032 & 1.01065 & 0.74137  \\
5904 & 10.16129 & 3.15739 & 2.81089 & 3.60758 & 3.11436  \\ 
6205 &  6.36782 & 2.49305 & 2.07314 & 2.75719 & 2.21766  \\ 
6218 &  4.80000 & 1.84201 & 1.71409 & 2.11447 & 1.92698  \\ 
6254 &  5.86957 & 1.95273 & 1.81290 & 2.24338 & 2.03818  \\ 
6341 &  3.64583 & 2.04998 & 1.66786 & 2.24019 & 1.76554  \\ 
6779 &  2.57732 & 1.39733 & 1.13396 & 1.52498 & 1.19904  \\
6838 &  1.50746 & 0.68612 & 0.60120 & 0.77926 & 0.66092  \\
6934 &  2.62238 & 1.27816 & 0.96695 & 1.33509 & 0.99089  \\
7078 &  6.31731 & 3.08644 & 2.47106 & 3.34126 & 2.59605  \\ 
7089 &  6.83654 & 2.91484 & 2.33384 & 3.15573 & 2.45203  \\ 
     &          &         &         &         &          \\
Pal5 &  1.07527 & 0.17804 & 0.13165 & 0.18190 & 0.13319  \\
     &  1.07527 & 0.15419 & 0.11401 & 0.15754 & 0.11535  \\
     &  1.07527 & 0.08631 & 0.06382 & 0.08819 & 0.06457  \\
     &  1.07527 & 0.07089 & 0.05242 & 0.07243 & 0.05303  \\
\noalign{\smallskip}      
\hline                                                       
\end{tabular}                                                
\end{center}                                                 
\end{table}                                                  

An inspection of Table \ref{t:glo2} and Fig.\,\ref{f:glo2} discloses that GCs
are (partially) unbound, owing to (i) a less extended inner sphere, and (ii) a
massive outer sphere.   In this view, the presence of tidal tails or tidal
streams detected in e.g., NGC 362, NGC 6934, NGC  7078, NGC 7089, (Grillmair
et al. 1995); NGC 5904, NGC 6205, (Leon et al. 2000); and the above mentioned
NGC 5466 (Grillmair and Johnson 2006); Pal5 (Odenkirchen et al. 2002; Dehnen
et al. 2004); should be predicted for the whole amount of Galactic GCs,
regardless of tidal shoks during disk passages.

The formulation used for the
tidal radius relates to a necessary condition, implying GCs above the
straight line, $\eta=\gamma$, plotted in Figs.\,\ref{f:glob}-\ref{f:glo2},
could also be (partially) unbound.

If a similar amount of fractional dark matter with respect to the Galaxy,
$M_{j,{\rm C}}/M_{i,{\rm C}}=10$ in the case under discussion, is considered
for each GC and calculations are repeated, the position on the
$({\sf O}\gamma\eta)$ plane looks similar to Fig.\,\ref{f:glob}.   In other
words, a comparable amount of dark-to-visible mass ratio within GCs and the
Galaxy makes GCs lie inside or near the stability region, as expected from
Eq.\,(\ref{eq:muto}).

For a generic density profile, $g(R)\propto v^2(R)/R$ via Eqs.\,(\ref{eq:v2})
and (\ref{eq:g}) where, in general, $v(R)\propto R^\beta$ hence
$g(R)\propto R^{2\beta-1}$ locally.   Then no extremum point for the
gravitational acceleration is expected if $\beta>1/2$, $0\le R\le a_{\rm G}$,
while the contrary holds if $\beta>1/2$, $R\appgeq0$, and $\beta<1/2$,
$R\appleq a_{\rm G}$.   For comparison with the trend inferred from
observations, a sample $(N=37)$ of empirical rotation curves, $v(R)$, used in
a recent investigation (Marr 2015), is considered.   Incomplete rotation
curves lacking data for large $(R>10 {\rm kpc}; N_{10}=13)$ and short
$(R<2 {\rm kpc}; N_2=4)$ galactocentric distances, are excluded.   The
resulting subsample $(N_0=N-N_{10}-N_2=20)$ shows rotation curve slopes
substantially larger than unity for short galactocentric distances and nearly
flat for large galactocentric distances.   If inferred velocities are related
to stable circular orbits, the gravitational acceleration must necessarily
exhibit an extremum point (maximum absolute value), which excludes density profiles where no
extremum point is present, such as H, HI, HS.

\section{Conclusion}
\label{conc}

Homeoidally striated spherical-symmetric density profiles have been classified
with reference to
four basic matter distributions, namely (i) isodensity i.e. $\rho(r)=$ const;
(ii) isogravity i.e. $g(r)=$ const; (iii) isothermal i.e.
$v(r)=[GM(r)/r]^{1/2}=$ const; (iv) isomass i.e. $M(r)=$ const.   In
particular, the following density profiles have been included: pure power-law; 
cored power-law; polytropic; Plummer (1911), shortened as P; Hernquist (1990),
shortened as H; Begeman et al. (1991), shortened as I; Spano et al. (2008),
shortened as S; Burkert (1995), shortened as B; for one-component systems, and
a few combinations of the above mentioned ones, namely PI, HI, PS, HS, for
two-component systems.    Special effort has been devoted to the occurrence of
an extremum point in $g(r)$, where the gravitational attraction on a test
particle of unit mass attains the maximum absolute value.

Tidal effects on subsystems have been considered using a definition of tidal
radius which is related to a necessary condition.   More specifically, given
an embedded sphere (subsystem) within an embedding sphere (one-component or
two-component system), the embedded sphere attains the tidal radius when the
gravitational force from the embedding and the embedded sphere, on the point
placed on the boundary of the latter and lying between related centres, are
equal in strength but act on opposite sides.

With regard to one-component systems, density profiles currently used for
representing galaxies and/or dark matter haloes, among those listed in Table
\ref{t:gadp}, have been considered for the embedding sphere.   The trend of
the fractional instantaneous tidal radius, $\eta$, vs. the inverse fractional
distance, $\gamma$, has shown the stability region, $\gamma<\eta$, is far more
extended for density profiles (P, I, S, B) where the gravitational
acceleration attains a maximum absolute value, with respect to density
profiles (H) where the
gravitational acceleration is monotonically increasing in absolute value.   In
the former
alternative, the tidal radius takes place $(\gamma=\eta)$ for two distinct
configurations, while in the latter alternative the same holds for a single
configuration.
       
The above mentioned trend is exacerbated for considered two-component systems,
where PI and PS density profiles exhibit a restricted stability region for
embedded spheres, while HI and HS density profiles show no stability region.
The gravitational acceleration attains a maximum absolute value in the former
case,
while a monotonic behaviour occurs in the latter.   The main features of
two-component embedding spheres appear to depend on the inner subsystem, with
little effects arising from the outer. 

The location of 17 globular clusters studied in earlier attempts (Brosche et
al 1999; Caimmi and Secco 2003), for which the radius, the mass, and the
Galactocentric distance are known, has been shown on the $({\sf O}\gamma\eta)$
plane for assigned Galactic truncation radius, mass, concentration, and
density profiles among those listed in Table \ref{t:gadp}.   Restricting to
star and gas subsystem, sample globular clusters have been found to lie within
the stability
region, $\gamma<\eta$, except Pal5 and NGC 5466, which exhibit noticeable
tidal effects (e.g., Odenkirchen et al. 2002; Grillmair and Johnson 2006).
Taking nonbaryonic dark matter into consideration, it has been shown the
results remain
unchanged in the special case where the mass ratio between non baryonic and
baryonic matter within globular clusters and the Galaxy, at any distance from
the centre, attains a constant value.

On the other hand, all sample globular clusters have been found to lie outside
the stability
region if the Galaxy $(M_{i,{\rm G}}=10^{11}m_\odot, a_{i,{\rm G}}=30$ kpc)
is embedded within a nonbaryonic dark matter halo 
$(M_{j,{\rm G}}=10^{12}m_\odot, a_{j,{\rm G}}=100$ kpc), for PI, PS, HI, HS,
two-component density profiles, unless the dark-to-visible mass
ratio within single globular clusters is comparable to its counterpart within
the Galaxy.

A comparison of predicted rotation curves, $v(R)$, with a subsample $(N_0=20)$
of empirical rotation curves has shown the occurrence of an extremum point
(maximum absolute value) in the gravitational acceleration.   Accordingly, density profiles
where no extremum point takes place, such as H, HI, HS, have necessarily been
excluded.


\appendix
\section*{Appendix}

\section{Homeoidally striated ellipsoidal density profiles}
\label{a:elli}

Density profiles in dimensionless coordinates, expressed by
Eqs.\,(\ref{eq:rho})-(\ref{eq:csif}), may be extended to the case where the
isopycnic i.e. constant density surfaces are similar and similarly placed
ellipsoids, which implies the dimensionless radial coordinate, $\xi=r(\mu)/
r^\dagger(\mu)$, maintains unchanged on an arbitrary isopycnic surface, $r=
r(\mu)$, $\mu$ polar angle (e.g., Caimmi and Marmo 2003).   In the special
case of homogeneous
ellipsoids, the gravitational potential and force on points not outside the
ellipsoids are (e.g., Caimmi and Secco 1992):
\begin{lefteqnarray}
\label{eq:Vh}
&& {\cal V}(x_1,x_2,x_3)=\pi G\rho\sum_{r=1}^3A_r(a_r^2-x_r^2)~~; \\
\label{eq:Fh}
&& \frac{\partial{\cal V}}{\partial x_r}=-2\pi G\rho A_rx_r~~;
\end{lefteqnarray}
where $a_r$ are the semiaxes of the ellipsoid and $A_r$ are shape factors for
which the following inequalities hold (e.g., MacMillan 1930, Chap.\,II, \S33;
Caimmi 1996):
\begin{lefteqnarray}
\label{eq:Ai}
&& A_1\le A_2\le A_3~~;\qquad a_1\ge a_2\ge a_3~~; \\
\label{eq:Aai}
&& a_1A_1\le a_2A_2\le a_3A_3~~;\qquad a_1\ge a_2\ge a_3~~;
\end{lefteqnarray}
accordingly, the largest gravitational force on the boundary of a homogeneous
ellipsoid is exerted at the top of the minor axis, ${\sf P}^\mp\equiv(0,0,\mp
a_3)$.   Using Newton's theorem, the above result can be extended to a
generic inner ellipsoid with similar and similarly placed boundary.

\begin{figure*}[t]
\begin{center}
\includegraphics[scale=0.8]{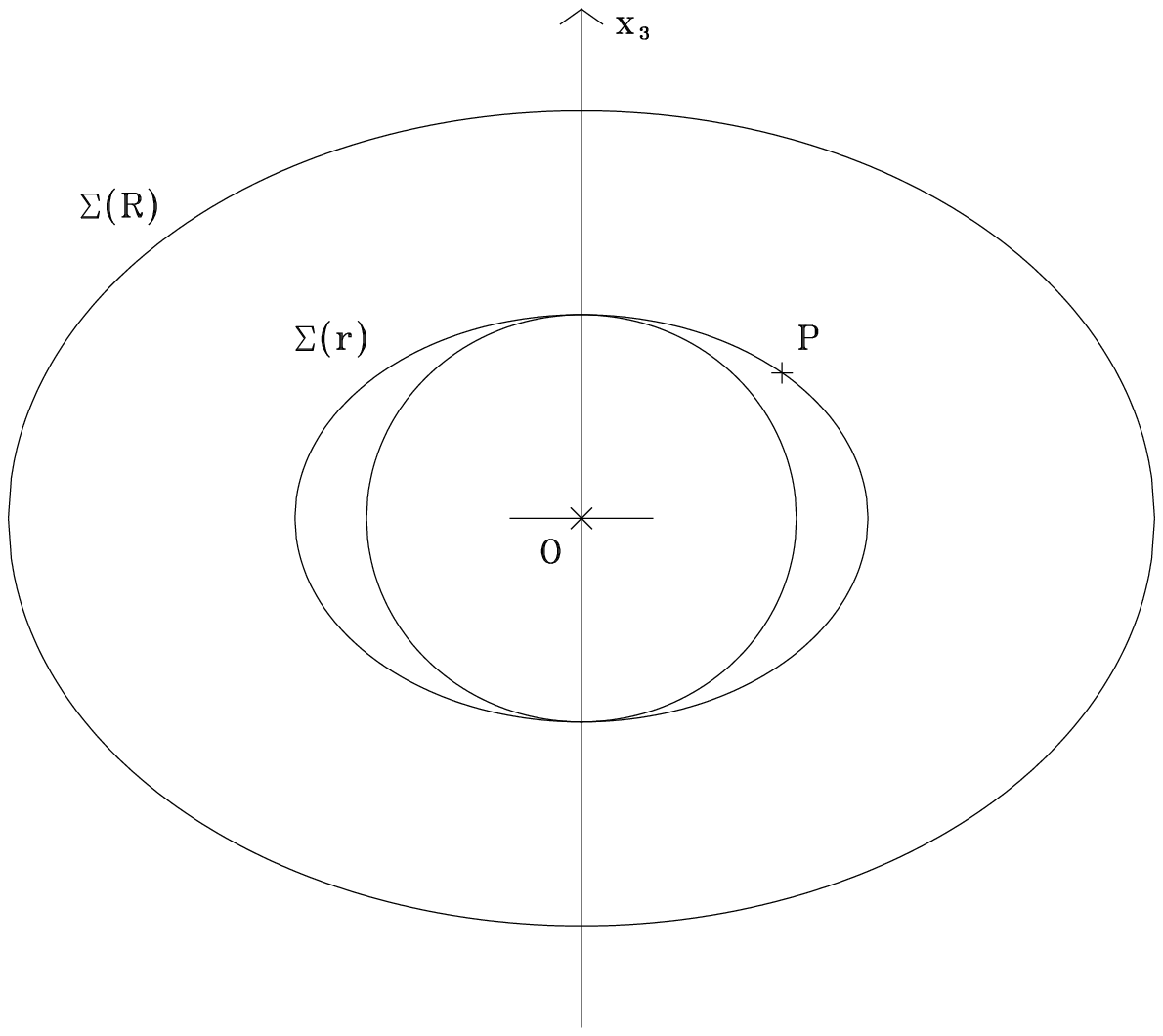}
\caption[OKND]{The attraction exerted by a homeoidally striated ellipsoid on
an internal or surface point, ${\sf P}\equiv(x_1,x_2,x_3)$, lies
between its
counterparts related to a homeoidally striated sphere and to a focaloidally
striated (with respect to the isopycnic surface passing through ${\sf P}$)
ellipsoid where, in both cases, isopycnic surfaces tangent at the minor axis
enclose an equal mass, $m(x_3)$.   See text for further details.   The outer
ellipsoid represents the boundary of the homeoidally striated ellipsoid,
$\Sigma(R)$.   The inner ellipsoid represents the isopycnic surface of the
homeoidally striated ellipsoid, $\Sigma(r)$, passing through ${\sf P}$, also
shown (cross).   The
sphere represents the isopycnic surface of the homeoidally striated sphere,
enclosing the same mass as $\Sigma(r)$.
The horizontal bar represents the focal ellipse with respect to the
focaloidally striated ellipsoid, enclosing
the same mass as the central isopycnic surface of the homeoidally striated
ellipsoid (saltire).   The coordinate axes, $x_1$ and $x_2$, are not drawn to
avoid confusion.}
\label{f:omco}
\end{center}
\end{figure*}

Turning to the general case of homeoidally striated ellipsoids, let ${\sf P}
\equiv(x_1,x_2,x_3)$ be a generic point within the ellipsoid, as represented
in Fig.\,\ref{f:omco}.

Owing to Newton's theorem, no attraction on ${\sf P}$ is exerted by the
homeoid bounded by the external isopycnic surface, $\Sigma(R)$, and the one
where ${\sf P}$ lies, $\Sigma(r)$.    Then only the ellipsoid bounded by
$\Sigma(r)$ has to be considered.

Let $\Sigma(r)$ together with all the enclosed isopycnic surfaces be
``compressed'' along $x_1$ and $x_2$ directions to attain a spherical shape
with radius equal to the minor axis of $\Sigma(r)$, as depicted in
Fig.\,\ref{f:omco}.   The result is a homeoidally
striated sphere with mass equal to the mass bounded by $\Sigma(r)$.   It is
suggested from geometrical considerations that the attraction exerted on
${\sf P}$ by the homeoidally striated sphere is larger than the attraction
exerted on  ${\sf P}$ by the homeoidally striated ellipsoid.

Let all the isopycnic surfaces within $\Sigma(r)$ be
``stretched'' along $x_1$ and $x_2$ directions, with the minor axis left
unchanged, to attain a confocal
ellipsoidal shape with respect to $\Sigma(r)$, as depicted in
Fig.\,\ref{f:omco}.   The result is a focaloidally
striated ellipsoid with mass equal to the mass bounded by $\Sigma(r)$ and
boundary $\Sigma(r)$.    Let $a_1^\pprime$, $a_2^\pprime$, $a_3^\pprime$, be
the semiaxes of a generic isopycnic surface within $\Sigma(r)$, and
$a_1^\prime$, $a_2^\prime$, $a_3^\prime$, the semiaxes of $\Sigma(r)$.   The
semiaxes of an ellipsoid internal and confocal to $\Sigma(r)$ are $c_r^\pprime
=\sqrt{(a_r^\prime)^2-\lambda}$, where $\lambda$ is determined by the boundary
condition, $c_3^\pprime=a_3^\pprime$, as $\lambda=(a_3^\prime)^2-
(a_3^\pprime)^2$.   Accordingly, the equatorial semiaxes of the confocal
ellipsoid are:
\begin{equation}
\label{eq:coho1}
(c_r^\pprime)^2=(a_r^\prime)^2-\lambda=(\epsilon_{r3}a_3^\prime)^2-
(a_3^\prime)^2+(a_3^\pprime)^2~~;
\end{equation}
which is equivalent to:
\begin{equation}
\label{eq:coho2}
(c_r^\pprime)^2=[(\epsilon_{r3})^2-1][(a_3^\prime)^2-(a_3^\pprime)^2]+
(a_r^\pprime)^2~~;
\end{equation}
where $\epsilon_{r3}=a_r/a_3=a_r^\prime/a_3^\prime=a_r^\pprime/a_3^\pprime$,
$r=1,2$, are the axis ratios of the isopycnic surface.   As $\epsilon_{r3}\ge
1$, $a_3^\prime\ge a_3^\pprime$, Eq.\,(\ref{eq:coho2}) discloses that, in
fact, the equatorial semiaxes of the confocal ellipsoid are stretched with
respect to the equatorial semiaxes of the related isopycnic surface,
$c_r^\pprime\ge a_r^\pprime$, $r=1,2$.

It is suggested from geometrical
considerations that the attraction exerted on ${\sf P}$ by the focaloidally
striated ellipsoid is lower than the attraction exerted on ${\sf P}$ by the
homeoidally striated ellipsoid.   Owing to MacLaurin's theorem, the attraction
exerted by a focaloidally striated ellipsoid on a surface point, ${\sf P}$,
equals the attraction exerted on the same point by a homogeneous ellipsoid
of equal mass and boundary.

A homeoidally striated ellipsoid may be conceived as a homogeneous ellipsoid
with equal boundary and density as at the surface isopycnic, to which is
superimposed an infinity of homogeneous ellipsoids bounded by isopycnic
surfaces, $\Sigma(r^\pprime)$, with infinitesimal density, $\diff\rho=\rho
(r^\pprime)-\rho(r^\pprime+\diff r)$.   Owing to MacLaurin's theorem, the
attraction exerted by the generic homogeneous ellipsoid on a surface point,
${\sf P}$, equals the attraction exerted by a homogeneous ellipsoid with equal
mass, confocal and confocally placed with respect to the one under
consideration, with ${\sf P}$ on its boundary.   The largest attraction
exerted on the boundary of a homogeneous ellipsoid is at the top of the minor
axis, which implies the largest attraction exerted on the boundary of a
homeoidally striated ellipsoid is also at the top of the minor axis.

For the homeoidally striated sphere, the gravitational attraction along the
minor axis, via Eq.\,(\ref{eq:Fh}) keeping in mind $A_r=2/3$ for spherical
configurations, reads:
\begin{equation}
\label{eq:Fgs}
\frac{\partial{\cal V}}{\partial x_3}=-\frac{Gm(x_3)}{x_3^2}~~;
\end{equation}
where $m(x_3)$ is the mass of the homeoidally striated ellipsoid enclosed
within the isopycnic surface with minor axis equal to $x_3$.

For the focaloidally striated ellipsoid, the gravitational attraction along
the minor axis, via Eq.\,(\ref{eq:Fh}) reads:
\begin{equation}
\label{eq:Fgc}
\frac{\partial{\cal V}}{\partial x_3}=-\frac32\frac{Gm(x_3)}{x_3^2}\epsilon_{31}
\epsilon_{32}A_3~~;
\end{equation}
which is different from its counterpart related to the homeoidally striated
sphere, Eq.\,(\ref{eq:Fgs}), by a shape factor, $(3/2)\epsilon_{31}
\epsilon_{32}A_3$, as expected.

Accordingly, the extremum point of the radial profile (along the minor axis)
of the attraction i.e. the gravitational acceleration is the same for both the
homeoidally striated sphere and the focaloidally striated ellipsoid, as the
two profiles are vertically shifted one with respect to the other, by an
amount equal to the shape factor.   Then it may safely be expected the
extremum point of the gravitational acceleration, related to the homeoidally
striated ellipsoid, is close to its counterpart related to the homeoidally
striated sphere and the focaloidally striated ellipsoid.

\section{Geometrical interpretation of the dimensionless scaling radius for
generalized power law density profiles}
\label{a:geme}

Dimensionless generalized power law density profiles, defined by
Eq.\,(\ref{eq:fgb}), in decimal logarithmic scale are expressed as:
\begin{equation}
\label{eq:gbl}
\log f=\log(C_\gamma+1)+\chi\log(C_\alpha+1)-\log(C_\gamma+\xi^\gamma)-\chi
\log(C_\alpha+\xi^\alpha)~~;
\end{equation}
where all parameters cannot be negative.

The logarithmic derivatives, $\diff^n\log f/\diff(\log\xi)^n$, can be
calculated as
\linebreak
$[\diff^n\log f/\diff\xi^n][\diff\xi/\diff\log\xi]^n=\xi^n
[\diff^n\log f/\diff\xi^n]$.    For the first, second, third, and fourth
derivative, the result is:
\begin{lefteqnarray}
\label{eq:dpl}
&& \frac{\diff\log f}{\diff\log\xi}=-\frac{\gamma\xi^\gamma}{C_\gamma+\xi^
\gamma}-\frac{\chi\alpha\xi^\alpha}{C_\alpha+\xi^\alpha}~~; \\
\label{eq:dsl}
&& \frac{\diff^2\log f}{\diff(\log\xi)^2}=-\frac{C_\gamma\gamma^2\xi^\gamma}
{(C_\gamma+\xi^\gamma)^2}-\frac{\chi C_\alpha\alpha^2\xi^\alpha}{(C_\alpha+
\xi^\alpha)^2}~~; \\
\label{eq:dtl}
&& \frac{\diff^3\log f}{\diff(\log\xi)^3}=-\frac{C_\gamma\gamma^3\xi^\gamma}
{(C_\gamma+\xi^\gamma)^3}(C_\gamma-\xi^\gamma)-\frac{\chi C_\alpha\alpha^3
\xi^\alpha}{(C_\alpha+\xi^\alpha)^3}(C_\alpha-\xi^\alpha)~~; \\
\label{eq:dql}
&& \frac{\diff^4\log f}{\diff(\log\xi)^4}=-C_\gamma\gamma^4\xi^\gamma\frac
{(C_\gamma-\xi^\gamma)^2-2C_\gamma\xi^\gamma}{(C_\gamma+\xi^\gamma)^4}-\chi
C_\alpha\alpha^4\xi^\alpha\frac{(C_\alpha-\xi^\alpha)^2}
{(C_\alpha+\xi^\alpha)^4}~~;\qquad
\end{lefteqnarray}
and the particularization to the dimensionless scaling radius, $\xi=1$,
$\log\xi=0$, yields:
\begin{lefteqnarray}
\label{eq:dpl1}
&& \left[\frac{\diff\log f}{\diff\log\xi}\right]_{\log\xi=0}=-\frac
{\gamma}{C_\gamma+1}-\frac{\chi\alpha}{C_\alpha+1}~~; \\
\label{eq:dsl1}
&& \left[\frac{\diff^2\log f}{\diff(\log\xi)^2}\right]_{\log\xi=0}=-\frac
{C_\gamma\gamma^2}{(C_\gamma+1)^2}-\frac{\chi C_\alpha
\alpha^2}{(C_\alpha+1)^2}~~; \\
\label{eq:dtl1}
&& \left[\frac{\diff^3\log f}{\diff(\log\xi)^3}\right]_{\log\xi=0}=-\frac
{C_\gamma\gamma^3(C_\gamma-1)}{(C_\gamma+1)^3}-\frac{\chi C_\alpha\alpha^3}
{(C_\alpha+1)^3}(C_\alpha-1)~~; \\
\label{eq:dql1}
&& \left[\frac{\diff^4\log f}{\diff(\log\xi)^4}\right]_{\log\xi=0}=
-\frac{C_\gamma\gamma^4[(C_\gamma-1)^2-2C_\gamma]}{(C_\gamma+1)^4}-
\frac{\chi C_\alpha\alpha^4}{(C_\alpha+1)^4}(C_\alpha-1)^2~~;\qquad
\end{lefteqnarray}
where Eq.\,(\ref{eq:dpl1}) coincides with (\ref{eq:pbg}), as $\chi\alpha=
\beta-\gamma$ by definition.

Turning to the general case and keeping in mind all parameters cannot be
negative, the first and the second derivative are always negative while the
third derivative is null provided the following relation:
\begin{equation}
\label{eq:dst0}
\frac{C_\gamma\gamma^3\xi^\gamma(C_\gamma-\xi^\gamma)}
{(C_\gamma+\xi^\gamma)^3}+\frac{\chi C_\alpha\alpha^3
\xi^\alpha(C_\alpha-\xi^\alpha)}{(C_\alpha+\xi^\alpha)^3}=0~~;
\end{equation}
is satisfied.   In particular, Eq.\,(\ref{eq:dst0}) holds at the dimensionless
scaling radius, $\xi=1$, for $C_\gamma=C_\alpha=1$ (B density profiles);
$C_\gamma=1, C_\alpha=0$ (Z density profiles); $C_\gamma=0, C_\alpha=1$
(Z density profiles).   Accordingly, the slope variation rate, $\diff^2\log f
/\diff(\log\xi)^2$, has an extremum point at $\xi=1$ where, in addition, the
fourth derivative turns out to be positive, which implies the slope variation
rate attains a minimum (a maximum in absolute value) at $\xi=1$.

\end{document}